\documentclass[times,twocolumn,final]{elsarticle}

\usepackage{cag}
\usepackage{framed,multirow}

\usepackage{amssymb}
\usepackage{amsmath}
\usepackage{latexsym}
\usepackage{booktabs}
\usepackage{CJKutf8} 
\usepackage{url}
\usepackage{xcolor}
\definecolor{newcolor}{rgb}{.8,.349,.1}

\usepackage{hyperref}

\usepackage{ulem}
\usepackage{graphicx}
\usepackage{animate}
\usepackage{multirow}
\usepackage{tabularx}
\usepackage{array}
\usepackage{subcaption}
\usepackage{float}    
\usepackage{balance}

\newcolumntype{Y}{>{\arraybackslash}X}

\begin{document}


\begin{frontmatter}

\title{SingVisio: Visual Analytics of Diffusion Model for Singing Voice Conversion}%
\author[1]{Liumeng Xue \fnref{fn1}}
\ead{xueliumeng@cuhk.edu.cn}
    
\author[1]{Chaoren Wang \fnref{fn1}}
\ead{chaorenwang@link.cuhk.edu.cn}
\fntext[fn1]{Equal contribution. This work is supported by NSFC (project no. 62376237 and 62302422), Shenzhen Science and Technology Program ZDSYS20230626091302006.} 

\author[1]{Mingxuan Wang}
\ead{mingxuanwang1@link.cuhk.edu.cn}

\author[1]{Xueyao Zhang}
\ead{xueyaozhang@link.cuhk.edu.cn}

\author[2]{Jun Han}
\ead{hanjun@ust.hk}

\author[1,3]{Zhizheng Wu}
\ead{wuzhizheng@cuhk.edu.cn}

\address[1]{The Chinese University of Hong Kong, Shenzhen, China}
\address[2]{The Hong Kong University of Science and Technology, China}
\address[3]{Shanghai AI Laboratory, Shanghai, China}

\begin{abstract}
In this study, we present SingVisio, an interactive visual analysis system that aims to explain the diffusion model used in singing voice conversion. SingVisio provides a visual display of the generation process in diffusion models, showcasing the step-by-step denoising of the noisy spectrum and its transformation into a clean spectrum that captures the desired singer's timbre. The system also facilitates side-by-side comparisons of different conditions, such as source content, melody, and target timbre, highlighting the impact of these conditions on the diffusion generation process and resulting conversions. Through comparative and comprehensive evaluations, SingVisio demonstrates its effectiveness in terms of system design, functionality, explainability, and user-friendliness. It offers users of various backgrounds valuable learning experiences and insights into the diffusion model for singing voice conversion.

\end{abstract}

\end{frontmatter}


\section{Introduction}
Deep generative models have become increasingly prevalent in a myriad of data generation tasks, ranging from image generation to audio generation. Among these, diffusion-based generative models have emerged as a cutting-edge research focus and the go-to methodology for such applications~\cite{yang2023diffusion}. In the field of computer vision, diffusion models have gained significant popularity~\cite{zhang2023text,xing2023survey}, particularly in applications such as text-to-image synthesis~\cite{Xu_2023_ICCV,rombach2021highresolution,rombach2021highresolution}, video generation~\cite{xing2023survey} and editing~\cite{ceylan2023pix2video}. In the audio community, there have been extensive studies of diffusion models in waveform synthesis~\cite{chen2020wavegrad,kong2020diffwave}, sound effects generation~\cite{liu2023audioldm,huang2023make}, speech generation~\cite{popov2021grad,naturalspeech2}, and music generation~\cite{diffsinger,schneider2023mo}. Given their wide-ranging utility and impressive performance, there is a burgeoning curiosity and necessity to unravel the intricacies of the diffusion process underpinning these generative tasks. However, the complexity of the involved Markov chains and their complex mathematical formulations pose a significant hurdle to novices in the field. In recent years, visual and interactive methodologies have proven instrumental in deciphering the structures and working mechanisms in various deep-learning models ~\cite{ganlab,lee2023diffusion,park2024explaining}. This insight has spurred us to develop interactive visual tools aimed at broader audiences, facilitating a deeper comprehension of diffusion-based generative models. The paper represents an attempt to demystify the diffusion-based generative paradigm.

Owing to its notable capabilities, the diffusion-based generative model has quickly risen as a formidable contender in singing voice conversion (SVC). This advanced technique effectively alters one singer's voice to another's, meticulously preserving the song's original content and melody, as investigated in the studies~\cite{diffsvc,zhang2023leveraging,lu2024comosvc}. When juxtaposed with other generative models, such as Generative Adversarial Networks (GANs)~\cite{goodfellow2020generative} and Variational Auto-Encoders (VAEs)~\cite{kingma2013vae}, diffusion-based models resolve the issue of unsatisfactory audio quality via incrementally introducing noise into the data and iteratively learning to eliminate noise. Due to iterative noising and denoising processes in synthesizing high-quality data, comparing the changes in the diffusion process step-by-step is essential to learn about the diffusion model. The current pedagogical approaches for beginners\footnote{In the context of this study, ``beginners'' are defined as individuals who have less than one year of experience in both the field of machine learning and the field of music and singing processing. This group primarily consists of users who are new to both the technical aspects of machine learning and the specific applications in music and singing processing. We expect the beginners' main focus to be on gaining fundamental knowledge about the diffusion model applied in the SVC in this study.} learning about diffusion-based models is overly dependent on textual explanations and mathematical descriptions\footnote{\url{https://theaisummer.com/diffusion-models/}}. This traditional learning method is neither intuitive nor efficient, often causing beginners to lose track among complex formulas without the ability to directly view and compare results at each step. Moreover, understanding the impact of various conditions—such as the source voice's content, melody, and the target singer's unique timbre—on the generation process is crucial for experts to identify challenging samples for SVC and make informed decisions to enhance SVC performance. Currently, comparing the effects of different conditions on SVC results is both time-consuming and cumbersome. Researchers must generate and save each feature, such as Mel spectrograms and audio files, and then repeatedly open and compare these across various steps.



Methods involving visualization and exploratory interaction are less common, as evidenced by examples such as ~\cite{karagiannakos2022diffusionmodels} and ~\cite{assemblyai_diffusion_2024}, which do not offer users an immersive understanding of the diffusion process. This highlights an urgent demand for comprehensive, interactive, and visually intuitive tools designed for diffusion-based generative models to fill this gap. 
In this paper, we propose SingVisio, a visual analytics system designed to interactively explain diffusion models in SVC. To maintain anonymity during the review process, the code will be made publicly available upon the paper's acceptance. SingVisio offers both a basic version to help beginners grasp the basic concepts of diffusion models, and an advanced version for experts by providing an efficient tool to further investigate diffusion-based SVC.
For visual representation, we extract Mel spectrograms and F0 contours from audio. 
Additionally, we demystify the diffusion process by extracting and rendering hidden features from different layers in the model over 1000 steps. 
Furthermore, we propose a novel interval clustering center sampling method, enabling users to flexibly specify the number of sample points and display the corresponding hidden features.

The contributions of this work can be summarized as follows:
\begin{itemize} 
    \item \textbf{A visual analytics system for understanding SVC.}
    To the best of our knowledge, this is the first system supporting the exploration, visualization, and comparison of the diffusion model within the context of SVC. It offers a versatile platform for comparing various aspects of the diffusion process, SVC modes, and evaluation metrics, allowing for a thorough exploration.

    \item \textbf{Novel interactive exploration approach to understanding diffusion-based SVC.}
    We have supported three core interactive exploration modes within our system: {\bf data-driven} exploration, which is steered by varying melodies, {\bf condition-driven} exploration that pivots on the specific inputs provided to the diffusion model, and {\bf evaluation-driven} exploration, which is based on the assessment metric. Also, we propose a novel interval clustering center sampling method to efficiently sample and display hidden features at specified steps.


    \item \textbf{A comparative and comprehensive evaluation of SingVisio.}
    We conducted a comparative and comprehensive evaluation of our system with the basic version and advanced version, including a case study involving two beginners, an expert study with two experts, and a formal user study encompassing both subjective and objective assessments for general users. Such evaluation shows the effectiveness of our system.
  
\end{itemize}

\section{Related Work}
\subsection{Singing Voice Conversion}
The early singing voice conversion research aims to design parametric statistical models such as HMM~\cite{parallel-svc-2009-HMM} or GMM~\cite{parallel-toda-2014,parallel-toda-2015} to learn the spectral features mapping of the parallel data. Since the parallel singing voice corpus is challenging to collect on a large scale, the non-parallel SVC~\cite{non-parallel-svc-facebook,non-parallel-svc-chenxin}, or recognition-synthesis SVC~\cite{self-supervised-vc}, has been popular in recent years, whose pipeline is displayed in Fig.~\ref{fig:svc-pipeline}. In the non-parallel SVC pipeline, the acoustic model conducts the feature conversion from source to target. It can be various types of generative models, including autoregressive models~\cite{non-parallel-svc-chenxin}, GAN-based models~\cite{fastsvc,zero-shot-roboust-svc-bgm}, VAE-based models~\cite{svc-technique,SoftvcVITS2023}, or Flow-based models~\cite{SoftvcVITS2023}. Besides, adopting a diffusion-based acoustic model is also promising for VC~\cite{diffvc,diff-hiervc} and SVC~\cite{diffsvc,zhang2023leveraging,lu2024comosvc}. Recently, more and more research has verified the strong performance of diffusion models in modeling audio areas~\cite{diffsinger,kong2020diffwave,naturalspeech2,audit}. 

Although the diffusion model has shown impressive quality and performance when applied to SVC, our understanding of its internal mechanisms is still limited. Firstly, the existing diffusion models are still based on black-box neural networks. Visualizing how it achieves singing voice conversion through step-by-step denoising would greatly deepen researchers' comprehension of the diffusion model's operating principles. Secondly, the SVC conditions, which serve as inputs to the diffusion model, are crucial factors influencing the final conversion results. However, we are still unclear about how different conditions affect the performance of the diffusion model. Motivated by that, this paper will conduct a systematic analysis of diffusion-based SVC under different diffusion steps and diverse SVC conditions, like varied sources and targets.

\subsection{Visual Analysis for Explainable AI }
E\textbf{X}plainable \textbf{A}rtificial \textbf{I}ntelligence (XAI)~\cite{arrieta2020explainable} has become increasingly important as machine learning models, especially deep learning models, grow in complexity and usage in critical applications~\cite{hohman2018visual}. Visual analysis tools have been developed to make these models more interpretable and trustworthy to users. CNN Explainer simplifies the understanding of Convolutional Neural Networks (CNNs) by visualizing their feature extraction process~\cite{wang2020cnn}. LSTMVis~\cite{LSTMVis} and DQNViz~\cite{DQNViz} offer insights into the decision-making processes of LSTM networks and Deep Q-Networks, respectively. M2Lens~\cite{M2lens} and CNNVis~\cite{CNNVis} are designed to dissect the intricate layers of CNNs, providing a detailed examination of filter activations and network architectures. AttentionViz focuses on the attention mechanisms in models, revealing how models prioritize different parts of the input data for decision-making~\cite {AttentionViz}.

Additionally, the interpretation of generative models through visualization addresses the challenge of understanding complex data generation processes. Adversarial-Playground~\cite{norton2017adversarial}, GANLab~\cite{ganlab} and GANViz~\cite{GANViz} are interactive tools for exploring and interpreting Generative Adversarial Networks (GANs). Research on analyzing the training processes of deep generative models uncovers the dynamics and stability issues inherent in these models. Further, DrugExplorer~\cite{DrugExplorer} exemplifies the application of visualization techniques in domain-specific areas. Recently, diffusion models have shown significant capabilities in generative tasks, and accordingly the visualization tool, aiming at making the diffusion process comprehensible to humans, is investigated~\cite{park2024explaining}. Besides, Diffusion Explainer concentrates on demystifying the stable diffusion process, offering an understanding of the transformation from text prompts into images~\cite{lee2023diffusion}. In our work, we design an interactive visual analysis system for the diffusion model applied in singing voice conversion. It illustrates how the noisy spectrum is gradually denoised under the influence of conditions, ultimately converting the spectrum to the target singer's timbre.

\section{Background: Diffusion-based Singing Voice Conversion}
\label{sec:background}

\begin{figure*}[t]
    \centering
    \includegraphics[width=0.65\textwidth]{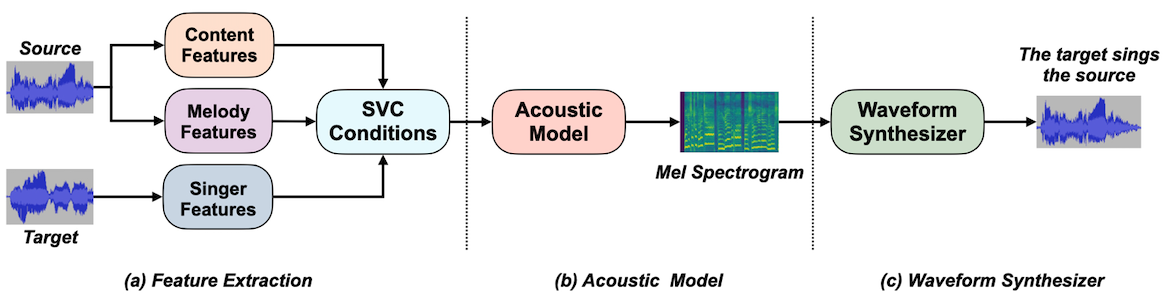} 
    \vspace{-10pt}
    \caption{The classic pipeline of SVC system, including three steps: (a) feature extraction that extracts content and melody features from the source and singer timbre from the target, (b) acoustic model mapping extracted features to acoustic features (e.g. Mel spectrogram), (c) waveform synthesizer reconstructing singing voice from the converted acoustic feature.  In this study, we use ``diffusion-based singing voice conversion" to refer that the acoustic model in the SVC is a diffusion model.
     \vspace{-29pt}
    }
    \label{fig:svc-pipeline}
\end{figure*}




SVC aims to transform the voice in a singing signal to match that of a target singer while preserving the original lyrics and melody~\cite{Huang2023TheSV}. The classic pipeline of SVC typically involves three steps, as shown in Fig.~\ref{fig:svc-pipeline}. (a) Feature extraction: extract content (i.e., lyrics) and melody features from the source singing voice and the timbre feature from the target singing voice. These features are then combined to form the conditions for SVC, which are fed into the following acoustic models. (b) Acoustic model: convert the source features to acoustic features (such as the Mel spectrogram) that match the target singer’s voice. (c) Waveform synthesizer: Reconstruct the singing voice waveform from the transformed acoustic features to produce the target singer timber while maintaining the source content.
In this study, the term `diffusion-based singing voice conversion' is used to denote that the acoustic model in the SVC system is a diffusion model. 

\subsection{Architecture and Workflow}

In this study, we select DiffWaveNetSVC~\cite{amphion,zhang2023leveraging} as the SVC's acoustic model to visualize and analyze. The internal module of the DiffWaveNetSVC is based on Bidirectional Non-Causal Dilated CNN~\cite{diffsvc,kong2020diffwave}, which is similar to WaveNet~\cite{Wavenet}. 

The architecture of DiffWaveNetSVC is shown in Fig.~\ref{fig:diffwavenetsvc} of \ref{app:svc_model}. It consists of multiple residual layers, within which it adopts Bidirectional Non-Causal Dilated CNN (``Bi-Dilated Conv" in Fig.~\ref{fig:diffwavenetsvc}) of \ref{app:svc_model} like~\cite{Wavenet,kong2020diffwave,diffsvc}. During training (i.e., the forward process of diffusion model), we extract the content, melody, and singer features from the same sample (which means the source and target in Fig.~\ref{fig:svc-pipeline} are the same) and add them to obtain the SVC conditions  $\mathbf{c}$. At the step $t \in [0, 1, 2, \cdots T]$, we sample a Gaussian noise $\mathbf{\epsilon}_t \sim N(\mathbf{0},~\mathbf{I})$ and obtain the noisy Mel spectrogram:
\begin{equation}
    \mathbf{y}_t = \sqrt{\alpha_t} \mathbf{y}_0 + \sqrt{1 - \alpha_t} \mathbf{\epsilon}_t,
\end{equation}
where $\alpha_t$ is the noise weight in diffusion model~\cite{ddpm}. And the training objective can be considered to predict the noise $\mathbf{\epsilon}_t$ using the neural network:
\begin{equation}
\begin{split}
    \hat{\mathbf{\epsilon}}_t &= \mathbf{DiffWaveNetSVC}(t, \mathbf{y}_t, \mathbf{c}), \\
    \mathcal{L}_t &= \mathbf{MSE}(\hat{\mathbf{\epsilon}}_t, {\mathbf{\epsilon}}_t),
\end{split}
\end{equation}
where $\mathbf{DiffWaveNetSVC}$ represents the whole encoder based on the residual layers and $\mathbf{MSE}$ means the mean squared error loss function.

During inference/conversion (the reverse process of diffusion model), given the source and target, we extract the content and melody features from the source, extract the singer features from the target, and add them as the SVC conditions $\mathbf{c}$. We feed a Gaussian noise $\hat{\mathbf{y}}_T \sim N(\mathbf{0},~\mathbf{I})$ to DiffWaveNetSVC and employ deep denoising implicit models~\cite{ddpm} with $T$ denoising steps to produce Mel spectrogram $\hat{\mathbf{y}_0}$.

\subsection{Implementation Details and Evaluation Metrics}
In this paper, we follow the Amphion's implementation~\cite{amphion}\footnote{\href{https://github.com/open-mmlab/Amphion/tree/main/egs/svc/MultipleContentsSVC}{https://github.com/open-mmlab/Amphion/tree/main/egs/svc/MultipleContentsSVC}} for DiffWaveNetSVC. Specifically, the layer number $N$ is 20, and the diffusion step number $T$ is 1000. Following Zhang et al.~\cite{zhang2023leveraging}, we adopt both Whisper~\cite{whisper} and ContentVec~\cite{contentvec} as the content features, we use Parselmouth\footnote{\href{https://parselmouth.readthedocs.io/en/stable/index.html}{https://parselmouth.readthedocs.io/en/stable/index.html}}~\cite{parselmouth} to extract F0 as the melody features, and we adopt look-up table to obtain the one-hot singer ID as the singer features. We utilize the DiffWaveNetSVC checkpoint of Zhang et al.~\cite{zhang2023leveraging} to conduct the inference, conversion, and visualization analysis, which is pre-trained on 83.1 hours of speech (111 singer) and 87.2 hours of singing data (96 singers). The detailed information about the dataset is described in~\ref{app:dataset}. For waveform synthesizer, we use the pre-trained Amphion Singing BigVGAN\footnote{\href{https://huggingface.co/amphion/BigVGAN_singing_bigdata}{https://huggingface.co/amphion/BigVGAN\_singing\_bigdata}} to produce waveform from Mel spectrogram.

Accurately and effectively assessing the results of SVC is significantly important~\cite{Huang2023TheSV}. Objective evaluation involves measuring performance at various aspects, such as spectrogram distortion, F0 modeling, intelligibility, and singer similarity.
To objectively evaluate synthesized samples, we adopt the evaluation methodology from Amphion~\cite{amphion}\footnote{\href{https://github.com/open-mmlab/Amphion/tree/main/egs/metrics}{https://github.com/open-mmlab/Amphion/tree/main/egs/metrics}} for our objective assessment. This includes metrics such as \textbf{Singer Similarity (Dembed) with Resemblyzer~\footnote{\url{https://github.com/resemble-ai/Resemblyzer}}}, \textbf{F0 Pearson Correlation Coefficient (F0CORR)}, \textbf{Fréchet Audio Distance (FAD)}, \textbf{F0 Root Mean Square Error (F0RMSE)},  and \textbf{Mel-cepstral Distortion (MCD)}. Detailed definitions of these metrics are provided in ~\ref{app:metircs}.

\section{Design Requirements}
\subsection{Requirement analysis}
Through a series of interviews with experts in audio signal processing and machine learning, we identified three critical tasks that our system needs to support for effective analysis and interpretation of the diffusion model for SVC.

\textbf{C1: In-Depth Temporal Dynamics Analysis of Diffusion Generation Process.}
Experts highlighted the importance of visualizing the temporal dynamics of the diffusion generation process in singing voice conversion. The objective is to create detailed visual representations that effectively trace the step-by-step evolution occurring in voice conversion at each diffusion stage. This involves visualizing the progression of various acoustic parameters, such as frequency components and harmonics that evolve over time. The visualizations are expected to provide users with an intuitive understanding of the voice transformation, highlighting the nuanced evolution from a noisy beginning to a structured and coherent output, thereby making the process more transparent and understandable to users without deep technical expertise in machine learning or signal processing.

\textbf{C2: Comprehensive Performance Metrics Evaluation in Singing Voice Synthesis.}
This task is centered on tracking evaluation metrics that gauge the quality of the converted voice at each step of the diffusion generation process. These metrics include pitch accuracy, timbre consistency, naturalness, and speech quality. The system should enable a detailed analysis of how each metric evolves with every diffusion step, offering insights into the conversion quality at different phases of the generation process. This comprehensive evaluation is pivotal in identifying aspects where the conversion achieves optimal quality or, conversely, where improvements are needed. This visual representation enhances the interpretability of the evaluation results and fosters insights into the underlying dynamics of the diffusion generation process.

\textbf{C3: Comparative Analysis of Different Source Singers, Songs, and Target Singers.}
Through visualization, we aim to systematically compare how different characteristics of source singers, such as vocal tone, pitch range, and singing style, influence the conversion outcome. This will help in identifying specific attributes of source singers that are more amenable to conversion. Additionally, the complexity and structure of the song itself are crucial variables. Songs with intricate melodic lines or complex rhythms might pose more significant challenges in conversion processes. Equally important is the analysis of the target singers' characteristics. The system should visualize how well the model adapts the source singer's voice to match the timbre of the target singers. This could lead to valuable insights, such as identifying particularly challenging source-target pairings or songs that consistently yield high-quality conversions. Such analysis is not only crucial for understanding the current model's performance but also for guiding future improvements and applications in singing voice conversion technology.

C1 and C2 tasks are related to fundamental knowledge of the diffusion model, which is crucial and beneficial for beginners to understand diffusion models. In contrast, C3 task focuses on exploring the impacts of different conditions on SVC, which is more suitable for experts or researchers seeking an in-depth understanding and analysis of diffusion-based SVC. Accordingly, we design SingVisio in two versions: a basic version and an advanced version. Both versions include C1 and C2 tasks. Additionally, the advanced version encompasses C3 task, catering to the needs of experts and researchers.


\begin{figure*}[h!]
    \centering
    \includegraphics[width=\textwidth]{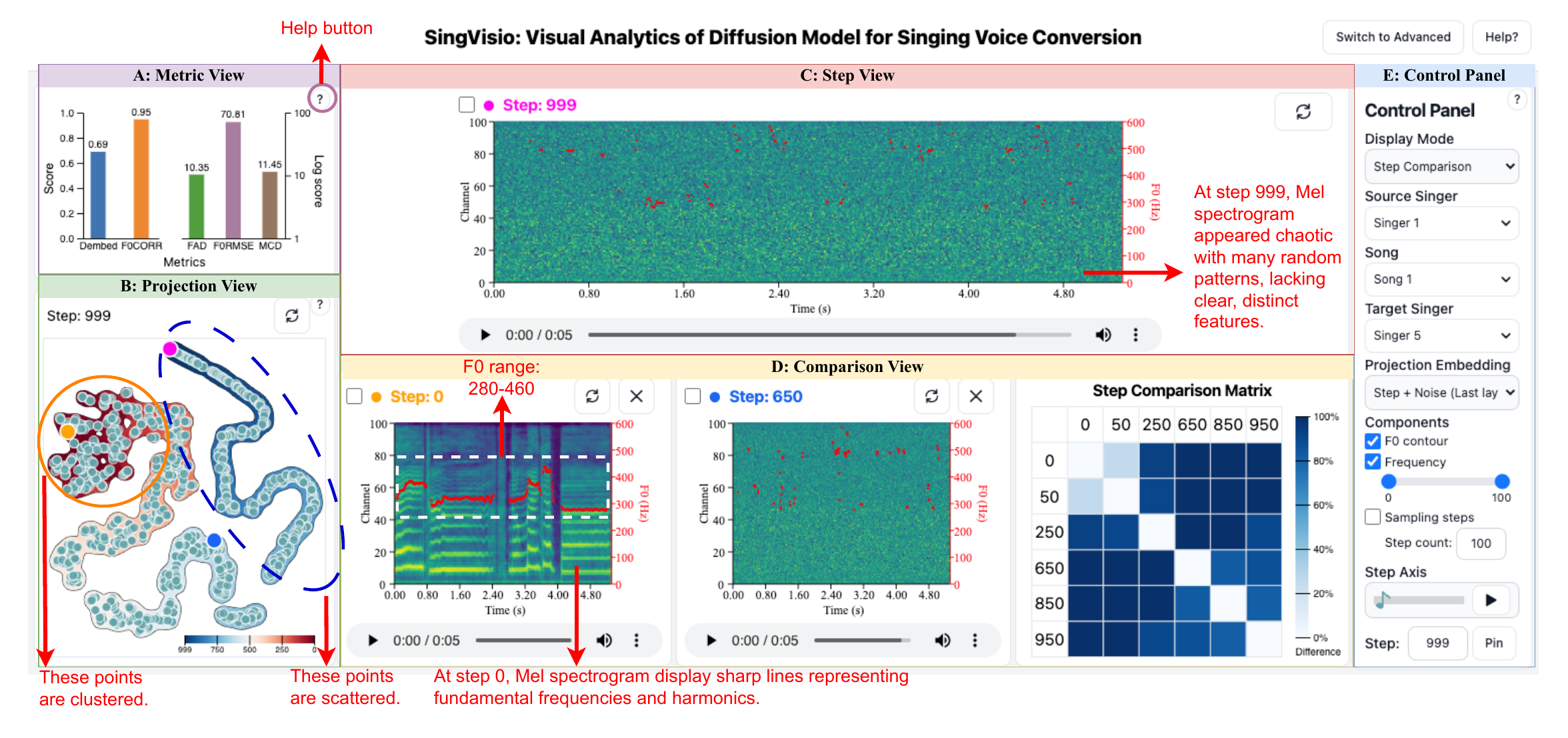}
     \vspace{-26pt}
    \caption{
    Visual system for diffusion-based singing voice conversion. The system consists of five views. 
    (A) \textbf{\textit{Metric View}} shows objective evaluation results on the singing voice conversion model, allowing users to interactively explore the performance trend along diffusion steps. (B) \textbf{\textit{Projection View}} aids users in tracking the data patterns of diffusion steps in the embedding space under different input conditions. (C) \textbf{\textit{Step View}} provides users with the visualization of Mel spectrogram and pitch contour at one diffusion step. (D) \textbf{\textit{Comparison View}} facilitates users to compare voice conversion results among different diffusion steps or singers. (E) \textbf{\textit{Control Panel}} enables users to select various comparison modes and choose different source and target singers to visually understand and analyze the model behavior. The red annotations provide explanations for the patterns or components. 
     \vspace{-19pt}
    }
    \label{fig:explainer_system}
\end{figure*}

\subsection{Analytical Tasks}\label{sec:analytical-tasks}
Our system is a visualization system designed specifically for diffusion-based SVC tasks. Diffusion-based SVC itself involves two aspects: in the realm of machine learning, it involves the diffusion generative model, and in the field of audio signal processing, it pertains to SVC. Therefore, the analytical tasks supported by our system can be divided into two major categories. In the aspect of machine learning, particularly in the diffusion model, to investigate the evolution and quality of the generated result from each step in the diffusion generation process, our system should support the following two tasks.

\textbf{T1: Step-wise Diffusion Generation Comparison.} Examining the generated result of each step in the diffusion generation process helps in understanding the model's behavior. Analyzing these early outputs can help us understand how the model initially handles noise. As each step incrementally adds detail and structure to the output, by inspecting intermediate steps, we can observe the step-by-step improvement in content quality. (C1)

\textbf{T2: Step-wise Metric Comparison.} As the diffusion steps progress, the generated content becomes clearer and more refined. Analyzing the objective evaluation metrics and their corresponding curves along the diffusion steps serves as a useful tool for assessing both the quantitative and qualitative aspects of the generated content. By tracking these metrics over the diffusion steps, we gain insights into how the model refines its output over time. (C2)

Regarding SVC, as described in Section~\ref{sec:background}, there are three factors (content, melody, singer timbre) that have a direct impact on the results of SVC and, therefore, should be considered during the conversion process. To explore the impact of different factors on the converted results, the system needs to provide support for the following three tasks.

\textbf{T3: Pair-wise SVC Comparison with Different \uline{Target Singers}.}  
Pair-wisely comparing SVC under two different conditions of the target singer at different diffusion steps. This task helps us to understand the impact of the timbre of the target singer that should be converted to the converted results of SVC, particularly in terms of singer similarity. (C1, C3)
    
\textbf{T4:  Pair-wise SVC Comparison with Different \uline{Source Singers}.}  
Pair-wisely comparing SVC under two different conditions of source singer at different diffusion steps. This task benefits us in exploring the impact of the melody of the source that should be kept on the converted results of SVC, particularly in terms of F0CORR, F0RMSE. (C1, C3)
    
\textbf{T5: Pair-wise SVC Comparison with Different \uline{Songs}.}  
Pair-wisely comparing SVC under two different conditions of the song at different diffusion steps. This task facilitates us to explore the impact of the content information conveyed in the song that should be maintained on the converted results of SVC, particularly in terms of MCD. (C1, C3)


\section{Explainer System}
Fig.~\ref{fig:explainer_system} shows the overview of the explainer system which consists of five components: \textbf{\textit{Control Panel}} allows users to modify mode and choose data for visual analysis; \textbf{\textit{Step View}} provides users with an overview of the diffusion generation process; \textbf{\textit{Comparison View}} makes it easy for users to compare converted results between different conditions; \textbf{\textit{Projection View}} helps users observe the trajectory of diffusion steps with or without conditions; \textbf{\textit{Metric View}} displays objective metrics evaluated on the diffusion-based SVC model, enabling users to interactively examine metric trends across diffusion steps.

\subsection{Control Panel}
\label{sec:control_panel}
The control panel consists of six components, including two drop-down boxes to enable users to select display mode and projection embedding, three checkboxes to select source singer, source song, and target singer, and a step controller to enable users to control the diffusion step.

    \textbf{Display Mode} We design five types of display modes, including Step Comparison, Source Singer Comparison, Song Comparison, Target Singer Comparison, and Metric Comparison. Users can click the drop-down box of ``Display Mode " to choose a specific model.
        \begin{itemize}
            \item \textbf{Step Comparison} This mode primarily focuses on step-wise comparing the diffusion steps in the generation process. It (1) provides an animation of random noise gradually refined for users to have an overview of the whole denoising process in \textbf{\textit{Step View}}, (2) enables users to adaptively select and compare the generated results from different diffusion steps in \textbf{\textit{Comparison View}}.

            \item \textbf{Metric Comparison} This mode presents five objective evaluation metric results of the diffusion-based SVC model represented by a bar chart. It (1) enables users to click on a specific metric bar and then the system filters out an example that gains the best on the corresponding metric and displays metric curves along diffusion steps in the \textbf{\textit{Comparison View}}, (3) enables users to hover over and slide the mouse along the step axis of the metric curve, and then system will display the values of that metric at different steps in the \textbf{\textit{Comparison View}} while synchronously showing the generated results at different steps in the \textbf{\textit{Step View}}.
            
            \item \textbf{Source Singer Comparison} This mode focuses on the pair-wise comparison of converting two different source singers' audio with the same song to the same target singer. It (1) allows users to select \uline{two different source singers}, a source song and a target singer, (2) provides the details (including Mel spectrogram, pitch contour, and audible audio) of the two source audio and the target audio in the \textbf{\textit{Comparison View}}, (3) presents two conversion animations wherein random noise undergoes gradual refinement to transform into the singing voice of the target singer in the \textbf{\textit{Step View}}. This mode is only available in the advanced version.
            
            \item \textbf{Song Comparison} This mode focuses on the pair-wise comparison of converting \uline{two different source audios} that are derived from the same singer but contain different songs to the same target singer. It (1) allows users to select a source singer, a target singer but two songs, (2) provides the details (including Mel spectrogram, pitch contour, and audible audio) of the two source singers' audio and the target singer's audio in the \textbf{\textit{Comparison View}}, (3) supplies two conversion animations illustrating the progressive refinement of random noise into the singing voice of the target in the \textbf{\textit{Step View}}. This mode is only available in the advanced version.

            \item \textbf{Target Singer Comparison} This mode focuses on the pair-wise comparison of converting the same source singing voice (also means the same song) to \uline{two different target singers}. It (1) enables users to select a source singer and a source song, but two target singers, (2) provides the details (including Mel spectrogram, pitch contour, and audible audio) of the source singer's audio, and two target singers' audio in the \textbf{\textit{Comparison View}}, (3) provides the two corresponding conversion animations of random noise gradually refined to the target singer singing voice in the \textbf{\textit{Step View}}. This mode is only available in the advanced version.
        \end{itemize}

    \textbf{Source Singer/Source Song/Target Singer} Three drop-down boxes offer users options for source singer, source song, and target singer.
    
    \textbf{Projection Embedding} A drop-down box to enable users to choose different projection embeddings from different layers. Then, the system displays 2D t-SNE visualization results of the high-dimensional diffusion steps in the \textbf{\textit{Projection View}}. Specifically, the projection embedding can be the diffusion steps, the combined embeddings of the step and noise, or step, noise and conditions. These embeddings can come from the first, middle, or final residual layer in the diffusion model, as illustrated in Fig.~\ref{fig:diffwavenetsvc} of \ref{app:svc_model}.

    \textbf{Components} Two checkboxes, labeled 'F0 contour' and 'Frequency,' allow users to control the display of these components in the Mel spectrogram. Additionally, the frequency bar lets users adjust the frequency range for display.
    
    \textbf{Step Controller} The Step Controller includes (1) a step slider to smoothly control the diffusion step, (2) a tool-tip to display or input a specific step number, and (3) a button named `Pin' that enables users to add a specific step's generated result in the \textbf{\textit{Comparison View}}.


\subsection{Step View}
This view enables users to visualize the whole generation process of diffusion in the context of SVC tasks, which means users can observe how the spectral characteristics change over time as noise is subsequently removed, leading to the desired SVC. Specifically, it can be observed that the Mel spectrogram transitions from being completely noisy to gradually becoming clearer, and the fundamental frequency curve also transforms from scattered points into a smooth curve. The audio also undergoes a process of gradual optimization from being pure noise to having improved sound quality and intelligibility. 

The control panel, mentioned earlier, allows users to interact with the diffusion process by smoothly sliding the step slider. Users can adjust the diffusion time step to observe the intermediate results of the generation process, enlarge the Mel spectrogram to observe detailed information through a brush operation, and restore it back to the original Mel spectrogram using the refresh button in the top right corner in the \textbf{\textit{Step View}}.

In the \textit{Step Comparison} and \textit{Condition Comparison} modes, the content presented in the \textbf{\textit{Step View}} is slightly different. In the \textit{Step Comparison} mode, we focus on comparing and analyzing the converted results from different steps, so only one diffusion process animation is displayed in the view. While, in the \textit{Condition Comparison} mode, the main objective is to compare the conversion results under different conditions, e.g., source singer, song, and target singer. At this time, the \textbf{\textit{Step View}} shows pair-wise diffusion process animations for two different conditions.

\begin{figure*}[ht!]
    \centering
    \includegraphics[width=0.9\textwidth]{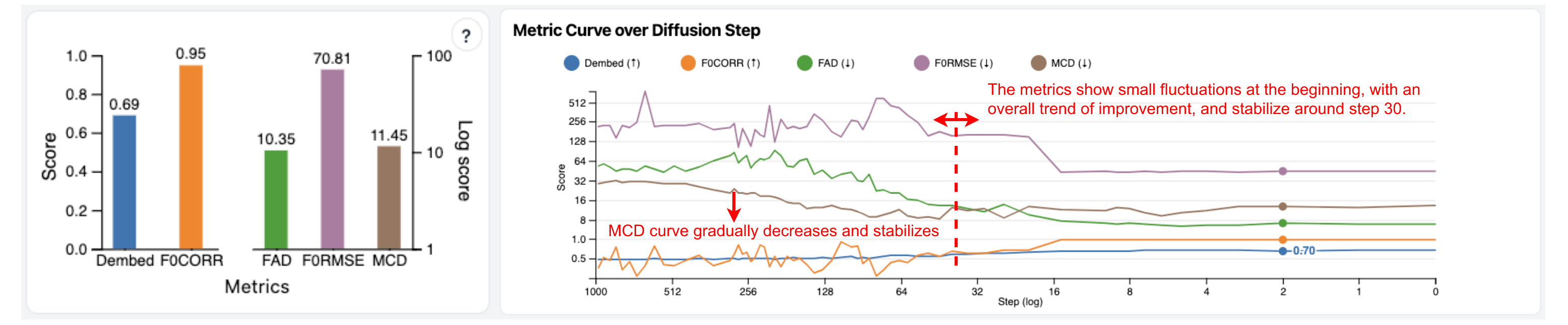}  
      \vspace{-8pt}
     \caption{The left part is the \textbf{\textit{Metric View}} with MCD metric selected. The right part is the corresponding ``Metric Curve over Diffusion Step" for the best-performing sample on the MCD metric. The red annotation in the right part explains the tendencies of metric curves.
    }
    \label{fig: metric_curve }
     \vspace{-19pt}
\end{figure*}

\subsection{Comparison View}

To facilitate a more convenient and detailed observation of the intermediate results generated by the diffusion model, we introduce a \textbf{\textit{Comparison View}}. Moreover, the comparison view differs between the basic and advanced versions. In the basic version, the comparison view initially displays a step comparison matrix, highlighting differences in Mel spectrograms between pairs of steps in the diffusion model, as shown in Fig.~\ref{fig:explainer_system}. Darker colors in the step comparison matrix indicate larger differences, while lighter colors represent smaller ones. Users can add specific steps to the matrix using the pin feature in the control panel or by clicking data points in the projection view. By clicking on the comparison matrix, Mel spectrograms and audio of the corresponding two steps can be displayed in the comparison view for detailed comparison. In the advanced version, we directly display three Mel spectrograms from three steps by default. Besides, users can select any step to replace the displayed three steps. It enables users to compare differences among three steps, broadening the scope of comparison.


It is noted that along with the Mel spectrogram, the corresponding audible audio, and fundamental frequency (F0) contour are also displayed in the Comparison View. All the information related to a clip of audio forms a basic block referred to as the ``basic display unit'', as shown in the below two Mel spectrograms in the comparison view in Fig.~\ref{fig:explainer_system} On this basic display unit, we can observe the range of the F0 and the pattern of the F0 contour. Through the brush operation, we can synchronously magnify all Mel spectrograms illustrated in this view, thus enabling a more detailed comparison and examination of the spectral differences. When there is more than one basic display unit, users can select the checkboxes in the top left corner of any two basic display units. The page will then pop up the visualization of the difference in the Mel spectrogram between these two basic display units, 
allowing for a clearer and more convenient comparison. Specifically, the differences are represented by colors. Warmer colors like reds and oranges signify larger differences, while cooler colors like blues and greens represent smaller differences between the two selected Mel spectrograms. This visualization aids in identifying which parts of the Mel spectrogram are significantly refined during the step-by-step generation process, highlighting areas that may require further investigation by algorithm researchers.

Furthermore, the components displayed in the \textbf{\textit{Comparison View}} differ between the \textit{Step Comparison Mode} and the \textit{Condition Comparison Mode}. The \textit{Step Comparison Mode} is primarily used to compare the results of different diffusion steps. In this mode, the \textbf{\textit{Comparison View}} will display basic display units from three different steps, based on the steps selected by the user. The relative position of different basic display units (corresponding to different steps) can be directly adjusted by dragging. On the other hand, the \textit{Condition Comparison Mode} sports a similar layout but mainly compares the results of SVC under different conditions. In this mode, the \textbf{\textit{Comparison View}} primarily displays the basic display units corresponding to different audios of the source and target singers selected by users. Additionally, in \textit{Metric Comparison Mode}, the \textbf{\textit{Comparison View}} illustrates the metric curve over the diffusion step, which is described in Section~\ref{sec:metric_view}.


\subsection{Projection View}\label{sec:projection-view}

High-dimensional hidden features can be challenging to interpret directly. t-SNE reduces the dimensionality by projecting the hidden feature embedding into a lower-dimensional space, allowing researchers to gain insights into the intricate structure and relationships within the high-dimensional space. By projecting high-dimensional step embeddings in the diffusion model into a lower-dimensional space, t-SNE reveals patterns and trajectories of the diffusion steps, enabling a visual exploration of the dynamic evolution of the diffusion process. Consequently, we design \textbf{\textit{Projection View}} to present the two-dimensional space obtained by projecting high-dimensional diffusion step embeddings (i.e., the step features in Fig.~\ref{fig:diffwavenetsvc} of \ref{app:svc_model}), as shown in Fig.~\ref{fig:explainer_system}. Each point represents a diffusion step, and all 1000 diffusion steps together form a trajectory in space. The boundary of this trajectory is highlighted with a gradient color scheme ranging from blue to red, reflecting the progression of the generation process. Users can hover their mouse over the points and slide to inspect the step trajectory. While sliding the mouse, users can simultaneously observe the SVC results transition from a coarse state to a fine state in \textbf{\textit{Step View}}. 
By scrolling the mouse wheel, they can zoom in or out on the points in the space to explore the distribution of the data points. By clicking on a specific step point, a basic display unit corresponding to the step will be added into the \textbf{\textit{Comparison View}}. 

As described in Section~\ref{sec:control_panel}, the drop-down menu of projection embedding in the control panel provides multiple projection embedding sources, including not only the vanilla diffusion step but also the combination of the diffusion step with noise and condition, as indicated by the red solid dots in Fig.~\ref{fig:diffwavenetsvc} of \ref{app:svc_model}. By examining the projection embedding results of combining diffusion step with noise and condition, users can compare the differences in diffusion step trajectories under different condition scenarios. Additionally, we propose a novel sampling strategy called interval clustering center sampling. The specific steps are as follows: (a) Set the number of steps to be sampled, denoted as $S$. (b) Divide the total interval into T/S sub-intervals, each sampling one sample. (c) Perform k-means clustering on all samples within each sub-interval to find the central point, and then calculate the distance from all samples to this center, and finally select the sample closest to the center as the representative sample. The clustering approach ensures that each sub-interval selection considers global temporal embedding information. Furthermore, by clustering and selecting the step closest to the center, the chosen steps are highly representative.

\subsection{Metric View}
\label{sec:metric_view}

\textbf{\textit{Metric View}} is designed to show the overall objective metrics evaluated on the model. The five metrics, including Dembed, F0CORR, FAD, F0RMSE, and MCD, are divided into two groups based on whether the values of the metrics are positively or negatively correlated with model performance and drawn in histograms. Here, the labels of the x-axis denote different metrics, and the labels of the y-axis are scores (the higher the better) and log scores (the lower the better). Each bar in the histogram is labeled with the calculated result corresponding to the metric. In the top right corner of this view, there is a button represented by a question mark. When users click this button, a tip box will appear providing descriptions of the definitions of each metric.


The \textbf{\textit{Metric View}} represents an average of all samples within the testing data pool, providing a comprehensive overview of the model performance with five objective metrics. Upon hovering over a particular metric, the system automatically identifies and selects the best-performing sample. This selection triggers a detailed visualization of the diffusion step for that sample within the \textbf{\textit{Projection View}}. Additionally, for the chosen sample, the system dynamically computes and displays evaluation metrics, which are then used to plot the ``Metric Curve over Diffusion Step" in \textbf{\textit{Comparison View}}, as shown in Fig.~\ref{fig: metric_curve }. 


At the top of the curve, five legends denoted as five different metrics are present with distinct colors. The x-axis shows different steps ranging from 999 to 0 as diffusion generates data, and the y-axis displays scores for evaluation metrics. The user can check the specific metric value for each step by hovering on the curve. Also, the step preview will update as the cursor moves on the curve. The interactive feature of \textbf{\textit{Metric View}} allows users to not only see aggregate metric performance but also delve into the 
variation trend of metrics with diffusion step during the diffusion generation process.

\subsection{Implementation Details}
The web application is designed to provide an interactive and user-friendly interface for visualizing spectrogram differences. It uses D3.js and TailwindCSS for the front-end, ensuring a clean and dynamic user interface. Specifically, D3.js handles the visualization, allowing for detailed and interactive spectrogram comparisons. TailwindCSS ensures a responsive and aesthetic design, enhancing user experience. The back-end is powered by Flask and Gunicorn, enabling efficient dynamic step sampling and efficient data retrieval with multiple workers. Specifically, Flask serves as the core framework, managing API requests and data processing. Gunicorn operates as the WSGI HTTP server, providing concurrency through multiple workers for fast data retrieval and processing. This architecture ensures that the application is both robust and scalable, capable of handling real-time spectrogram analysis and visualization efficiently.

Mel spectrogram and Fundamental Frequency  Contour (F0 contour) are extracted from the audio using a signal processing algorithm. Mel spectrogram is a 2D representation with the dimensions of Time*Channel, where the time axis captures the progression of the audio signal over time, and the channel axis represents the frequency components or Mel bins, providing a comprehensive view of the signal's spectral content. The Mel spectrogram is color-coded to indicate the intensity or magnitude of different frequencies over time. Bright colors, such as yellow and red, represent high energy or the presence of specific frequencies, while darker colors represent lower energy or the absence of those frequencies.
F0 contour is a key concept in the fields of speech processing and music analysis, especially in the study of prosody, intonation, and melody. It refers to the variation in the pitch of a voice over time. The fundamental frequency, or F0, is the lowest frequency of a periodic waveform and determines the pitch of the sound, which is one of the primary auditory attributes used to distinguish different sounds in speech and music. This contour line may be drawn as a continuous curve that rises and falls to depict changes in F0. The F0 contour line is colored red in this work, to distinguish it from the Mel spectrogram.


\section{Case Study}
We invited two beginners in machine learning and signal processing, E1 and E2, to participate in a case study to verify whether the system could make the model interpretable and help beginner users understand the working mechanism of the diffusion model applied in SVC tasks.



E1 focused on the step view, observing the transition of the Mel spectrogram from noisy to clean. Initially, the Mel spectrogram appeared \textbf{chaotic with many random patterns, lacking clear, distinct features}. As the process continued from step 999 to step 0, the spectrogram \textbf{gradually became clearer, displaying sharp lines representing fundamental frequencies and harmonics}. Correspondingly, \textbf{the initial audio sounded indistinct and lacked clarity, presenting hissing and other unwanted sounds}. Eventually, \textbf{the vocals became well-defined and easy to discern, with almost no unwanted sounds or interference}.
\textit{E1 commented that this dynamic display intuitively demonstrated the entire process of SVC, making the generation process more interpretable and comprehensible.} Moreover, \textit{E1 mentioned that listening to the voice transition from one blurred timbre to another clear timbre was quite fascinating.}


E2 mainly interacted with the system by dragging the step axis to control the diffusion reverse step, observing the differences in the generated results at various steps. E2 also focused on the metric view. E2 clicked on the help button shaped like a question mark in the top right corner of the metric view (as shown in Fig.~\ref{fig:explainer_system}) to learn about the definitions of metrics and their correlation with model performance. E2 then clicked on the MCD metric bar, prompting the system to show five Metric Curves over Diffusion Steps in the Comparison View. E2 moved the mouse over the MCD metric curve and the system displayed the corresponding MCD value,  Mel spectrogram and audio of the corresponding step. Additionally, E2 listened to the corresponding audio at different steps, providing an audible perception of the changes. E2 observed that all metric values changed from the starting point to a gradually stabilizing endpoint throughout the diffusion process. \textit{E2 mentioned that this was the first time they directly observed the fluctuations of metrics throughout the diffusion process.
E2 described the system as a comprehensive and user-friendly visualization tool for diffusion models in SVC tasks that allows for both an overview and a detailed study. 
}


\section{Expert Study}
We invite two domain experts (E3 and E4) to participate in an expert study to evaluate the system based on its usability and effectiveness. They were not involved in the system design process, nor did they participate in the user study and case study. E3 is a researcher who has been engaged in machine learning and voice conversion research for more than 3 years. E4 is also a researcher primarily focusing on SVC and is strongly interested in XAI.


\textbf{System Usefulness}
Both experts acknowledged SingVisio as a valuable tool for validating domain knowledge. They observed that the system clearly demonstrates each step's results during the data generation process in the diffusion model.
Specifically, in a Mel spectrogram, noise appears as random speckles or fuzzy areas. Early in the reverse diffusion process (step 999), the spectrogram has high noise levels because the model is just starting to refine the audio. By step 50, the noise decreases, resulting in a cleaner spectrogram. This indicates successful noise reduction and improved audio quality. Harmonic structures, seen as horizontal lines and spectral patterns show the distribution of energy across frequencies. 
\textbf{Visual Designs and Interactions}
From the t-SNE visualization of projection embedding (such as step embedding) in the projection view, experts observe a distinct pattern transitioning from a decentralized to a more centralized structure (as illustrated in Fig~\ref{fig:explainer_system}). In the reverse process of a diffusion model, each step builds on the output of the previous step to remove noise. Viewed as an optimization problem, each step minimizes the difference between the original data and the current estimate. As this process progresses, the latent representations increasingly resemble the original data points, causing them to appear more clustered in the t-SNE plot.

\textbf{Insight and Inspiration}
Both domain experts believe the system provides valuable insights. They observed the transformation of the Mel spectrogram from noise to a clear signal during the diffusion generation process in SVC. It was found that when the target singer's F0 is low and dense, more steps are required for the signal to become clear, indicating greater difficulty in converting to such target singers. Frequent and dense F0 changes increase modeling complexity, necessitating more steps to accurately generate these variations while avoiding distortion and maintaining harmonic structure consistency. Fine-tuning model parameters for low and dense F0 cases can yield better results. Additionally, increasing the quantity and diversity of such data can enhance model robustness and generalization capability.

Additionally, E4 noted that the metric comparison perspective reveals the limitations of existing objective metrics used in SVC. For example, in a 1000-step diffusion model, almost all metric curves approach convergence within about the last 30 steps, showing no significant improvement beyond that point. However, from the step comparison perspective, we can see (and hear) a substantial difference in sound quality between the generated results at step 30 and step 0, indicating areas for further enhancement.
This observation suggests that while numerical metrics may indicate stabilization, the perceptual quality of audio continues to improve significantly in the final stages of the diffusion process. E4 emphasized that this discrepancy highlights a critical gap in current evaluation methods, as metrics like MCD, FAD, and F0RMSE may not fully capture the nuanced improvements audible to human listeners. To address this, E4 suggested developing new, perceptually aligned metrics that better reflect auditory differences observed during the final diffusion steps.


\section{Evaluation}
This section details the evaluation approach and the results of SingVisio in both basic and advanced versions.
The evaluation is carried out through structured user studies, designed to assess both objective understanding and subjective experiences of the users.


\subsection{User Study Set-up}

\textbf{Participants}
We recruited 23 participants (P1-P23) from audio, music, and speech processing laboratories. They included beginners new to the field, doctoral students with 1-2 years of experience, and postdoctoral researchers with 4-6 years of experience. This mix of participants ensured a broad perspective on the system's performance across different user groups. Additionally, their research interests centered on audio, music, and speech processing. While they had limited knowledge of visual analysis, they showed great interest in the SingVisio system as it allowed them to interactively visualize their research content, e.g., audio, Mel spectrograms, generative models.



\textbf{Questionnaire}
The questionnaire was designed to capture both objective and subjective aspects of the SingVisio. The questions are designed following ContextWing~\cite{zhao2023contextwing} and also considering the specific features of our own system.

\begin{itemize}
    \item \textbf{Objective Questions} \ In the part of objective questions, to evaluate the effectiveness of both the basic and advanced versions, two sets of questionnaires were designed. These questions are directly related to the five analytical tasks (T1-T5) previously detailed in Section~\ref{sec:analytical-tasks}.
    \begin{itemize}
        \item \textbf{Objective Questions (Basic Version)} (OB1-OB8) For participants in the basic version, the study was conducted as a comparative analysis, where users were divided into two subgroups. One group engaged with SingVisio, while the other group utilized the traditional tutorial method to learn about diffusion-based SVC models, with the same dataset (audio files, Mel spectrograms with F0 visualized, metric data spreadsheet) of every step provided, to answer the same set of questions. This comparative approach allowed us to directly assess the efficiency of SingVisio in helping beginners grasp concepts compared with conventional learning methods.
        \item \textbf{Objective Questions (Advanced Version)} (OA1-OA15) The questionnaire for the advanced version is designed with users with more experience or specialized in audio, music or speech processing, aiming to evaluate the effectiveness of the system in aiding field experts in facilitating a more sophisticated analysis and understanding.
    \end{itemize}
    \item \textbf{Common Subjective Questions} (S1-S16) Both versions of the questionnaire included a shared set of subjective questions, which are intended to capture the users’ perceptions, satisfaction, and any qualitative feedback regarding their experience. The questions are designed following ContextWing~\cite{zhao2023contextwing} in evaluating the system around four key aspects, including explainability, analysis functionality, design effectiveness, and usability. These aspects were selected based on the recommendations by Rossi et al.~\cite{rossi2018evaluation}, ensuring a comprehensive evaluation framework that aligns with established user experience principles. The questions are rated using a 5-point Likert scale, ranging from 1 (strongly disagree) to 5 (strongly agree). 
\end{itemize}
By employing this dual-version structure, our questionnaire not only provides insights into the specific utilities of each version of SingVisio but also allows for a nuanced analysis of its educational impact compared to traditional methods. This methodology supports a robust evaluation of the system’s effectiveness across a spectrum of users, from novices to advanced practitioners.



\textbf{Procedure}
We first divided the participants by their experience in the field of audio, music, or speech processing in terms of years ($<$1yr, basic group, $>=$1yr, advanced SingVisio group), the basic group is further split evenly and randomly into two sub-groups, basic SingVisio group and tutorial group. Out of the 23 participants, the advanced group consisted of 10 individuals. The basic group included 13 individuals, with 7 in the basic SingVisio group and 6 in the tutorial group. Both advanced and basic SingVisio groups were first oriented with a comprehensive introduction to the SingVisio system, familiarizing the participants with the functions and capabilities of SingVisio. The tutorial group was oriented with a tutorial session to learn about necessary knowledge about diffusion-based SVC models and basic concepts like F0, Mel spectrograms and metric definitions. Following the initial setup, participants proceeded to complete the online questionnaire. Those in the SingVisio groups answered the questions while actively using the SingVisio system, configured as specified in the questionnaire. Conversely, participants in the tutorial group answered the questions using a tutorial handout, whilst having access to the same dataset as the SingVisio groups. This dataset included audio files, Mel spectrograms with visualized F0, and a spreadsheet detailing metric data for each step of the process. After both objective and subjective queries were completed, an optional feedback section was provided for any additional comments or suggestions. 


\subsection{Results and Analysis}
The completion time for the basic tutorial group and basic SingVisio group on the basic version was approximately 94.31 (\(\sigma=78.23\)) minutes and 48.65 (\(\sigma=21.93\))  minutes, respectively. Additionally, the completion time for the advanced SingVisio group was about 40.54 (\(\sigma=26.62\)) minutes. 
It is noted that the completion time for the user study includes not only the time taken to answer the questionnaire but also the time spent familiarizing with the SingVisio system or tutorial. Additionally, the study was conducted in an uncontrolled environment, where participants used their own computers, resulting in potential distractions that could have affected the completion times. Even though removing the extreme outlier, the completion time of the tutorial group  (\(\mu =68.13\), \(\sigma=50.13\)) is greater than that of the other two groups. It indicates that the visual and interactive approach in the SingVisio system is more conducive to completing the questionnaire, thereby resulting in shorter completion times.

The average accuracies for the tutorial group and basic SingVisio group were 71.73\% and 82.14\%, respectively, and the average accuracy for this group was approximately 91.77\%. The results indicate that using the SingVisio system requires significantly less time to complete the user study compared to the traditional tutorial method. Meanwhile, SingVisio effectively aids beginners in understanding diffusion-based SVC more efficiently. In contrast, the traditional tutorial method involves manually finding and comparing audio or Mel spectrograms from thousands of files. SingVisio simplifies this by allowing the dynamic display of audio and Mel spectrograms at specified steps, thereby improving efficiency. 
Furthermore, the higher accuracy and reduced completion time observed among users of the advanced SingVisio group can be attributed to their professional background in signal processing. With at least two years of experience, these users are better equipped to efficiently navigate the system and effectively extract relevant information, contributing to their overall performance.


\vspace{-10pt}
\begin{figure}[th!]
    \centering
    \includegraphics[width=0.9\linewidth]{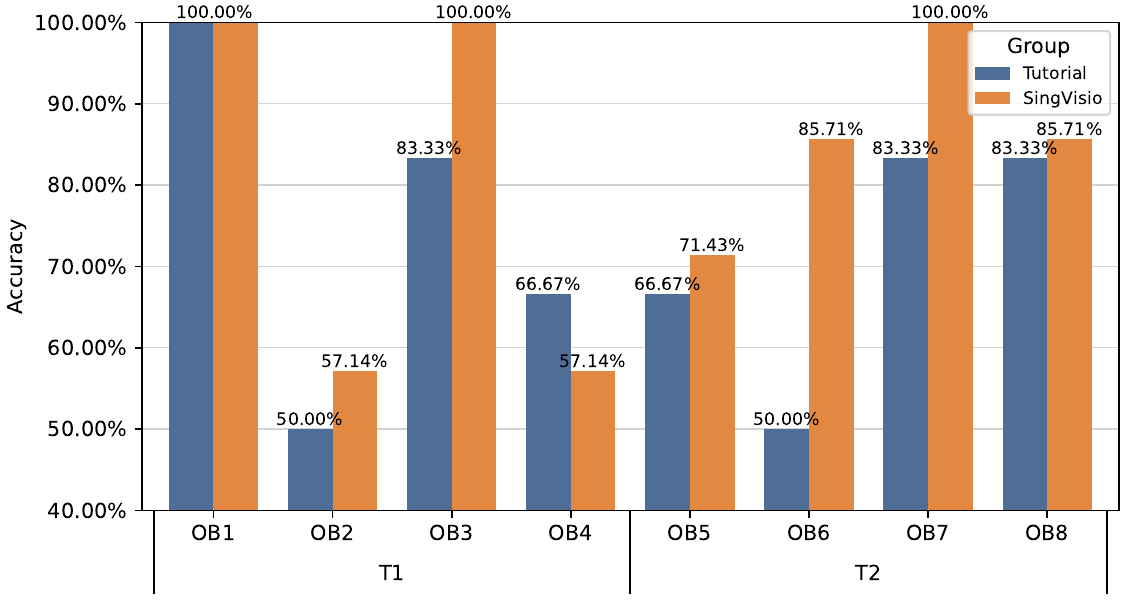}
    \vspace{-12pt}
    \caption{Accuracy of objective questionnaires on the basic version, including tutorial group and basic SingVisio group. The questions designed for the basic version are related to analysis tasks T1 and T2 as described in Section 4.2.}
    \label{fig:obj_bsc}
    \vspace{-25pt}
\end{figure}

\subsubsection{Objective Evaluation for basic version}
The accuracy of the objective questions (OB1-OB8) for the basic version is shown in Figure~\ref{fig:obj_bsc}, including the tutorial group and SingVisio group. Overall, all questions from the SingVisio group obtained higher accuracy than those from the tutorial group except for a question (OB4). The detailed questions and results are described as follows.

\textbf{Step-wise Diffusion Generation Comparison (T1). } 
We designed four questions (OB1-OB4) related to the diffusion generation process in SVC for the basic version.
The objective results shown in Fig.~\ref{fig:obj_bsc} show that the accuracy of OB1 from both groups were 100\%, indicating that the system's capability for users to learn about the diffusion generation process is on par with the tutorial. The accuracies of OB2 and OB3 from the SingVisio group were higher than those from the Tutorial group, indicating that the SingVisio system allows users to clearly observe the F0 range of the audio (as shown in the annotation in Fig.~\ref{fig:explainer_system}). While the tutorial group achieved slightly higher accuracy than the basic SingVisio group on OB4, further analysis provided insight into this discrepancy. The tutorial group benefited from a t-SNE visualization example with handwritten digit recognition, which clearly demonstrated clustering. In contrast, SingVisio's t-SNE pattern (as shown in the right bottom part of Fig.~\ref{fig:explainer_system} ) in the diffusion generation process, while forming clusters, was less apparent. This greater clarity in the tutorial's visual representation likely led to higher accuracy for this question in the tutorial group.


\textbf{Step-wise Metric Comparison (T2). }
For task T2, we designed four questions (OB5-OB8).
Comparing the accuracy rates for OB5-OB8, we found that the SingVisio group consistently outperformed the Tutorial group, with the most significant gains in OB6, followed by OB7, OB5, and OB8. OB6 and OB7, which focus on the overall trend of the metric curve (as annotated in Fig~\ref{fig: metric_curve }), showed that SingVisio's interactive and intuitive display of the complete curve is more effective than the tutorial's method of using Excel sheets to deduce trends. OB5 and OB8 involve understanding the relationship between metrics and model performance. SingVisio's helpful tool-tips explaining terms or concepts aid in better understanding, resulting in higher accuracy rates for SingVisio.

\vspace{-10pt}
\begin{figure}[h!]
    \centering
    \includegraphics[width=0.9\linewidth]{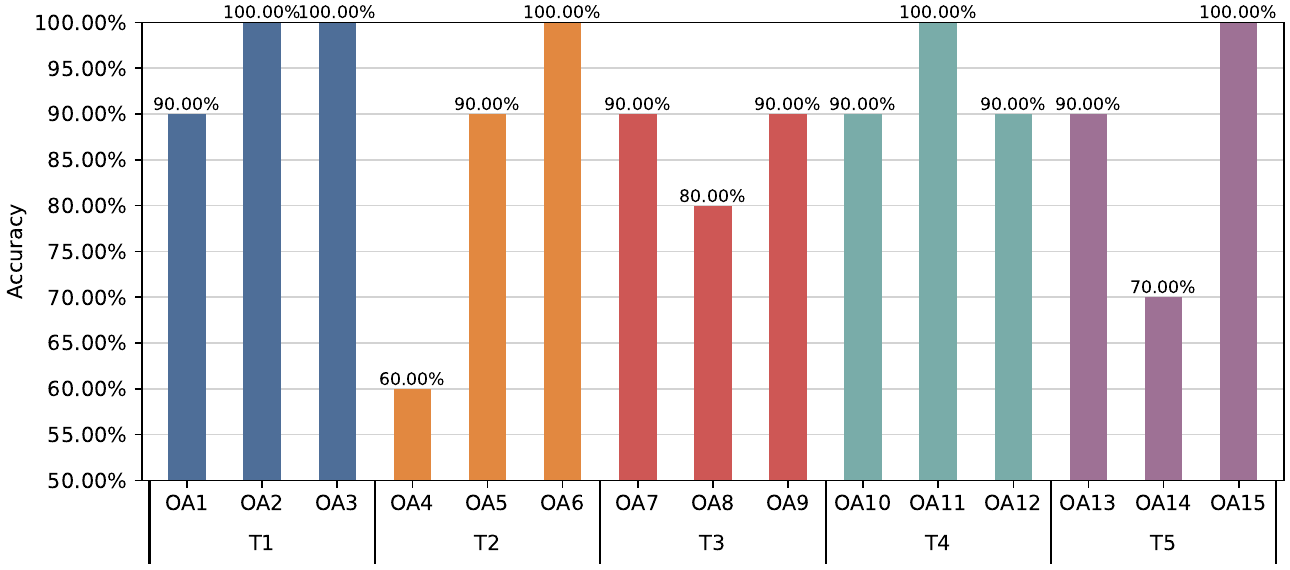}
    \vspace{-12pt}
    \caption{Accuracy of objective questionnaires on advanced version. The questions designed for the advanced version are related to all analysis tasks T1-T5, as described in Section 4.2.}
    \label{fig:obj_adv}
     \vspace{-27pt}
\end{figure}

\subsubsection{Objective Evaluation for advanced version}
The result of the objective evaluation for the advanced version is presented in Fig.~\ref{fig:obj_adv}. The data reveal that the majority of questions were answered with accuracies between 90\% and 100\%, with only three questions falling below this threshold. This suggests that SingVisio effectively supports researchers in answering queries pertinent to diffusion models and SVC. Considering the proficiency of advanced users in these subjects, the questions designed for this version (T1 and T2) were intentionally made more complex than those in the basic version. These 15 objective questions, designated OA1 to OA15, are described as follows.

\textbf{Step-wise Diffusion Generation Comparison (T1).}
Questions OA1-OA3 related to T1 are designed for the advanced version.
From the result shown in Fig.~\ref{fig:obj_adv} of the three questions relating to T1, OA1 achieved 90\% accuracy, while both OA2 and OA3 achieved 100\% accuracy. This indicates the effectiveness of SingVisio in helping users acquire knowledge about diffusion generation and understand the corresponding influence of the Mel spectrogram and F0 contour.

\begin{figure*}[h!]
    \centering
    \includegraphics[width=0.55\textwidth]{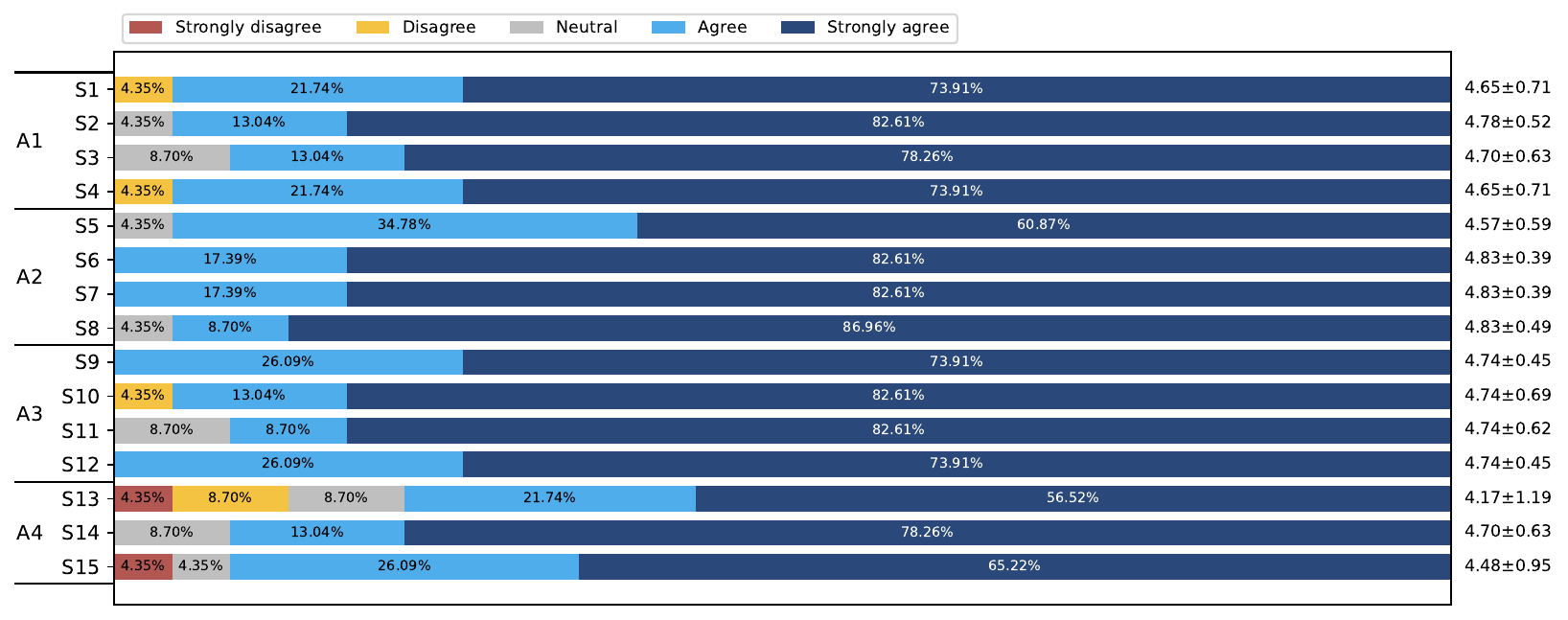}
    \vspace{-10pt}
    \caption{Rating scores of subjective questionnaires. A1-A4 represents the four aspects, including explainability, functionality, effectiveness, and usability. S1-S15 denotes 15 subjective questions. The rightmost value shows the mean ± standard deviation for each question.} 
    \label{fig:sbj}
     \vspace{-19pt}
\end{figure*}

\textbf{Step-wise Metric Comparison (T2).}
Three questions related to T2 are designed for the advanced version (OA4-OA6).
The ranking in the accuracy of the three questions related to T2, OA6, OA5 and OA4, were 100\%, 90\% and 60\%, respectively. The score for OA4 was relatively low. After consulting several users, we found that the descriptions "metrics show improvement" and "metrics show degradation" referred to whether the metric itself improves or degrades during the diffusion generation process, not whether its value increases or decreases. This misunderstanding led to a lower accuracy rate for OA4. In contrast, OA6 received a 100\% accuracy rate, indicating that the provided projection embedding is interpretable and allows users to recognize patterns.


\textbf{Pair-wise SVC Comparison with Different \uline{Target Singers} (T3).}
Regarding T3, there are questions (OA7-OA9).
Among these questions, OA7 and OA9 both achieved 90\% accuracy, while OA8 achieved 80\% accuracy. OA7 pertains to the timbre of the singing voice. SingVisio provides both a visual Mel spectrogram and audible audio. In SingVisio, users can select the specified singer via the control panel and listen to the corresponding audio, allowing flexible and efficient analysis while comparing it with the Mel spectrogram. OA9 involves analyzing the difficulty of converting the same source to different target singers. The 90\% accuracy indicates that SingVisio effectively helps users determine which conditions in SVC are easy and which are challenging.

\textbf{Pair-wise SVC Comparison with Different \uline{Source Singers} (T4).}
T4 aims to analyze and understand SVC under different source conditions. For T4, three questions (OA10-OA12) were designed.
From the results, we can find that all questions have high accuracy. Specifically, OA11 gets 100\% accuracy, and OA10 and OA12 obtain 90\% accuracy. OA11 involves analyzing whether two singers have different singing styles. This can be observed from the Mel spectrogram, where the harmonic patterns in density and position are noticeably different, and from the audio, where the differences in singing styles can be heard. OA10 pertains to analyzing the F0 (fundamental frequency) of two singers. This can be determined by observing the red-marked F0 contour in the Mel spectrogram. OA12 involves analyzing the duration and fundamental frequency of the conversion results. This information can also be obtained from both the Mel spectrogram and the audio. These results demonstrate that SingVisio provides an effective and flexible tool that offers both audible and visual insights, enabling users to gain comprehensive information from various perspectives.



\textbf{Pair-wise SVC Comparison with Different \uline{Source Songs} (T5).}
For T5, three questions were designed(OA13-OA15). 
The accuracy rankings for these questions are OA15, OA13, and OA14, with scores of 100\%, 90\%, and 70\% respectively. OA15 involves comparing the projection embedding patterns of two conversion processes. The projection view shows similar trajectories for the hidden features in both conversions, indicating that SingVisio's projection view is highly effective for analyzing hidden features in diffusion generation.

OA13 and OA14 pertain to the timbre and singing content of two conversion results. Although these questions should ideally have no errors, user inquiries revealed that users often assume the source in SVC includes both content and melody information. Our data includes the same singer performing different songs, which is why our control panel has separate settings for source singer and song. Users mistakenly assumed the source singer included the target content. For future studies, we will avoid ambiguous terms and clearly explain the conditions and questions.

\subsubsection{Subjective Evaluation}
We conducted subjective evaluations of four aspects (A1-A4), including explainability, functionality, effectiveness, and usability. Overall, the subjection evaluation comprises 15 subjective questions (S1-S15), each scored on a scale ranging from 1 to 5, i.e., strongly disagree (1), disagree (2), neutral (3), agree (4), and strongly agree (5). The assessment results yield an average score of 4.67 (\(\sigma=0.02\)).
Notably, \textbf{it achieved the highest score in analysis functionality (\(\mu = 4.76\), \(\sigma=0.13\)) and also performed well in effectiveness  (\(\mu = 4.74\), \(\sigma=0.0\))}. Detailed results for each dimension and question are presented in Fig.~\ref{fig:sbj}.

\textbf{Explainablility (A1)}
As shown in Fig.~\ref{fig:sbj}, the subjective assessment results across four dimensions indicate that explainability obtains the high score (\(\mu = 4.70\), \(\sigma=0.06\)), demonstrating the effectiveness of SingVisio in interpreting diffusion models and SVC.
Among the four questions designed to validate the explainability of SingVisio, S1-S4, S2 scored the highest (\(\mu = 4.78\), \(\sigma=0.52\)), indicating the system's effectiveness in aiding users to understand and explain metrics changing over the diffusion generation process. S1, S3, and S4 all received more than 4.45 scores, further confirming that our system provides a comprehensive understanding and explanation for the diffusion model in the context of the SVC task.



\textbf{Analysis Functionality (A2)}
The test results revealed that in the subjective assessment across four dimensions, the score in the analysis functionality dimension is the highest (\(\mu = 4.76\), \(\sigma=0.13\)). This dimension's subjective evaluation is designed to verify the system's support for analysis across tasks T1-T5 and includes four specific questions, S5-S8. Among these questions, S6, S7, and S8 scored the same highest score (\(\mu = 4.83\)), with all users agreeing or strongly agreeing that SingVisio supports analysis and comparison of generated results at different diffusion steps (T1) and the analysis of evaluation metrics (T2). S8, designed for the analysis tasks T3-T5, received agree and strongly agree from all but one neutral user, indicating that SingVisio effectively supports analysis for T3-T5.


\textbf{Visual Design Effectiveness (A3)}
To validate the effectiveness of our system design, we formulate four questions (S9-S12). 
All four questions scored about 4.74 points, indicating that all users agree or strongly agree that our views' design and the system's interactive design are effective. S10 and S11 received about 82.61\% strongly agree, demonstrating that  \textbf{\textit{Step View}} and \textbf{\textit{Comparison View}}, designed specifically for T1-T5, are effective.

\textbf{Usability (A4)}
To evaluate the usability of SingVisio, we design three related questions (S13-S15).
Participants in the user study included those with over three years of experience in machine learning and signal processing, some new to these fields, and others with a purely musical background. Over half of the users strongly believed in the user-friendly interface and ease of learning of SingVisio (S13 and S15). However, the presence of strong disagreement in S13 and S15 indicates that our system still requires improvements to enhance its user-friendliness and ease of use.  More than 78\% users strongly recommended SingVisio to others who could benefit from its use (S14). This demonstrates that our system is user-friendly for diverse users, regardless of their background. 


\section{Conclusion}
In this work, we introduce SingVisio, a visual analysis system designed to interactively explain the diffusion model for singing voice conversion. Specifically, SingVisio visually exhibits the step-wise generation process of diffusion models, illustrating the gradual denoising of the noisy spectrum, ultimately resulting in a clean spectrum that captures the target singer's timbre. The system also supports pairwise comparisons between different conditions, such as content and melody in source audio, and timbre from the target audio, revealing the impact of these conditions on the diffusion generation process and converted results. Comparative and comprehensive evaluations demonstrate that SingVisio is effective in terms of system design, functionalities, explainability, and usability. It provides diverse users with fresh learning experiences and valuable insights into the diffusion model for singing voice conversion.


\newpage
\bibliographystyle{cag-num-names}
\bibliography{main}

\begin{thebibliography}{61}
\providecommand{\natexlab}[1]{#1}
\providecommand{\url}[1]{\texttt{#1}}
\providecommand{\href}[2]{#2}
\providecommand{\path}[1]{#1}
\providecommand{\eprint}[1]{\href{http://arxiv.org/abs/#1}{\path{#1}}}
\providecommand{\DOIprefix}{doi:}
\providecommand{\ArXivprefix}{arXiv:}
\providecommand{\URLprefix}{URL: }
\providecommand{\Pubmedprefix}{pmid:}
\providecommand{\doi}[1]{\href{http://dx.doi.org/#1}{\path{#1}}}
\providecommand{\Pubmed}[1]{\href{pmid:#1}{\path{#1}}}
\providecommand{\BIBand}{and}
\providecommand{\bibinfo}[2]{#2}
\ifx\xfnm\undefined \def\xfnm[#1]{\unskip,\space#1}\fi
\bibitem[{Yang et~al.(2023)Yang, Zhang, Song, Hong, Xu, Zhao et~al.}]{yang2023diffusion}
\bibinfo{author}{Yang\xfnm[ L]}, \bibinfo{author}{Zhang\xfnm[ Z]}, \bibinfo{author}{Song\xfnm[ Y]}, \bibinfo{author}{Hong\xfnm[ S]}, \bibinfo{author}{Xu\xfnm[ R]}, \bibinfo{author}{Zhao\xfnm[ Y]}, et~al.
\newblock \bibinfo{title}{Diffusion models: A comprehensive survey of methods and applications}.
\newblock \bibinfo{journal}{ACM Computing Surveys} \bibinfo{year}{2023};\bibinfo{volume}{56}(\bibinfo{number}{4}):\bibinfo{pages}{1--39}.
\bibitem[{Zhang et~al.(2023{\natexlab{a}})Zhang, Zhang, Zhang and Kweon}]{zhang2023text}
\bibinfo{author}{Zhang\xfnm[ C]}, \bibinfo{author}{Zhang\xfnm[ C]}, \bibinfo{author}{Zhang\xfnm[ M]}, \bibinfo{author}{Kweon\xfnm[ IS]}.
\newblock \bibinfo{title}{Text-to-image diffusion model in generative {AI}: A survey}.
\newblock \bibinfo{journal}{arXiv preprint arXiv:230307909} \bibinfo{year}{2023}{\natexlab{a}};.
\bibitem[{Xing et~al.(2023)Xing, Feng, Chen, Dai, Hu, Xu et~al.}]{xing2023survey}
\bibinfo{author}{Xing\xfnm[ Z]}, \bibinfo{author}{Feng\xfnm[ Q]}, \bibinfo{author}{Chen\xfnm[ H]}, \bibinfo{author}{Dai\xfnm[ Q]}, \bibinfo{author}{Hu\xfnm[ H]}, \bibinfo{author}{Xu\xfnm[ H]}, et~al.
\newblock \bibinfo{title}{A survey on video diffusion models}.
\newblock \bibinfo{journal}{arXiv preprint arXiv:231010647} \bibinfo{year}{2023};.
\bibitem[{Xu et~al.(2023)Xu, Wang, Zhang, Wang and Shi}]{Xu_2023_ICCV}
\bibinfo{author}{Xu\xfnm[ X]}, \bibinfo{author}{Wang\xfnm[ Z]}, \bibinfo{author}{Zhang\xfnm[ G]}, \bibinfo{author}{Wang\xfnm[ K]}, \bibinfo{author}{Shi\xfnm[ H]}.
\newblock \bibinfo{title}{Versatile diffusion: Text, images and variations all in one diffusion model}.
\newblock In: \bibinfo{booktitle}{IEEE/CVF International Conference on Computer Vision}. \bibinfo{year}{2023}, p. \bibinfo{pages}{7754--7765}.
\bibitem[{Rombach et~al.(2022)Rombach, Blattmann, Lorenz, Esser and Ommer}]{rombach2021highresolution}
\bibinfo{author}{Rombach\xfnm[ R]}, \bibinfo{author}{Blattmann\xfnm[ A]}, \bibinfo{author}{Lorenz\xfnm[ D]}, \bibinfo{author}{Esser\xfnm[ P]}, \bibinfo{author}{Ommer\xfnm[ B]}.
\newblock \bibinfo{title}{High-resolution image synthesis with latent diffusion models}.
\newblock In: \bibinfo{booktitle}{Proceedings of the IEEE/CVF conference on computer vision and pattern recognition}. \bibinfo{year}{2022}, p. \bibinfo{pages}{10684--10695}.
\bibitem[{Ceylan et~al.(2023)Ceylan, Huang and Mitra}]{ceylan2023pix2video}
\bibinfo{author}{Ceylan\xfnm[ D]}, \bibinfo{author}{Huang\xfnm[ CHP]}, \bibinfo{author}{Mitra\xfnm[ NJ]}.
\newblock \bibinfo{title}{Pix2video: Video editing using image diffusion}.
\newblock In: \bibinfo{booktitle}{IEEE/CVF International Conference on Computer Vision}. \bibinfo{year}{2023}, p. \bibinfo{pages}{23206--23217}.
\bibitem[{Chen et~al.(2021)Chen, Sainath, Pang, Vaswani, Shazeer and Parmar}]{chen2020wavegrad}
\bibinfo{author}{Chen\xfnm[ B]}, \bibinfo{author}{Sainath\xfnm[ TN]}, \bibinfo{author}{Pang\xfnm[ RJ]}, \bibinfo{author}{Vaswani\xfnm[ A]}, \bibinfo{author}{Shazeer\xfnm[ N]}, \bibinfo{author}{Parmar\xfnm[ N]}.
\newblock \bibinfo{title}{{WaveGrad}: Estimating gradients for generative audio modeling}.
\newblock In: \bibinfo{booktitle}{International Conference on Learning Representations}. \bibinfo{year}{2021},.
\bibitem[{Kong et~al.(2020)Kong, Ping, Huang, Zhao and Catanzaro}]{kong2020diffwave}
\bibinfo{author}{Kong\xfnm[ Z]}, \bibinfo{author}{Ping\xfnm[ W]}, \bibinfo{author}{Huang\xfnm[ J]}, \bibinfo{author}{Zhao\xfnm[ K]}, \bibinfo{author}{Catanzaro\xfnm[ B]}.
\newblock \bibinfo{title}{{DiffWave}: A versatile diffusion model for audio synthesis}.
\newblock In: \bibinfo{booktitle}{International Conference on Learning Representations}. \bibinfo{year}{2020},.
\bibitem[{Liu et~al.(2023)Liu, Chen, Yuan, Mei, Liu, Mandic et~al.}]{liu2023audioldm}
\bibinfo{author}{Liu\xfnm[ H]}, \bibinfo{author}{Chen\xfnm[ Z]}, \bibinfo{author}{Yuan\xfnm[ Y]}, \bibinfo{author}{Mei\xfnm[ X]}, \bibinfo{author}{Liu\xfnm[ X]}, \bibinfo{author}{Mandic\xfnm[ D]}, et~al.
\newblock \bibinfo{title}{{AudioLDM}: Text-to-audio generation with latent diffusion models}.
\newblock \bibinfo{journal}{arXiv preprint arXiv:230112503} \bibinfo{year}{2023};.
\bibitem[{Huang et~al.(2023{\natexlab{a}})Huang, Ren, Huang, Yang, Ye, Zhang et~al.}]{huang2023make}
\bibinfo{author}{Huang\xfnm[ J]}, \bibinfo{author}{Ren\xfnm[ Y]}, \bibinfo{author}{Huang\xfnm[ R]}, \bibinfo{author}{Yang\xfnm[ D]}, \bibinfo{author}{Ye\xfnm[ Z]}, \bibinfo{author}{Zhang\xfnm[ C]}, et~al.
\newblock \bibinfo{title}{{Make-An-Audio 2}: Temporal-enhanced text-to-audio generation}.
\newblock \bibinfo{journal}{arXiv preprint arXiv:230518474} \bibinfo{year}{2023}{\natexlab{a}};.
\bibitem[{Popov et~al.(2021)Popov, Vovk, Gogoryan, Sadekova and Kudinov}]{popov2021grad}
\bibinfo{author}{Popov\xfnm[ V]}, \bibinfo{author}{Vovk\xfnm[ I]}, \bibinfo{author}{Gogoryan\xfnm[ V]}, \bibinfo{author}{Sadekova\xfnm[ T]}, \bibinfo{author}{Kudinov\xfnm[ M]}.
\newblock \bibinfo{title}{{Grad-TTS}: A diffusion probabilistic model for text-to-speech}.
\newblock In: \bibinfo{booktitle}{International Conference on Machine Learning}. \bibinfo{year}{2021}, p. \bibinfo{pages}{8599--8608}.
\bibitem[{Shen et~al.(2024)Shen, Ju, Tan, Liu, Leng, He et~al.}]{naturalspeech2}
\bibinfo{author}{Shen\xfnm[ K]}, \bibinfo{author}{Ju\xfnm[ Z]}, \bibinfo{author}{Tan\xfnm[ X]}, \bibinfo{author}{Liu\xfnm[ Y]}, \bibinfo{author}{Leng\xfnm[ Y]}, \bibinfo{author}{He\xfnm[ L]}, et~al.
\newblock \bibinfo{title}{{NaturalSpeech 2}: Latent diffusion models are natural and zero-shot speech and singing synthesizers}.
\newblock In: \bibinfo{booktitle}{International Conference on Learning Representations}. \bibinfo{year}{2024},.
\bibitem[{Liu et~al.(2022)Liu, Li, Ren, Chen and Zhao}]{diffsinger}
\bibinfo{author}{Liu\xfnm[ J]}, \bibinfo{author}{Li\xfnm[ C]}, \bibinfo{author}{Ren\xfnm[ Y]}, \bibinfo{author}{Chen\xfnm[ F]}, \bibinfo{author}{Zhao\xfnm[ Z]}.
\newblock \bibinfo{title}{{DiffSinger}: Singing voice synthesis via shallow diffusion mechanism}.
\newblock In: \bibinfo{booktitle}{AAAI Conference on Artificial Intelligence}. \bibinfo{year}{2022}, p. \bibinfo{pages}{11020--11028}.
\bibitem[{Schneider et~al.(2023)Schneider, Kamal, Jin and Sch{\"o}lkopf}]{schneider2023mo}
\bibinfo{author}{Schneider\xfnm[ F]}, \bibinfo{author}{Kamal\xfnm[ O]}, \bibinfo{author}{Jin\xfnm[ Z]}, \bibinfo{author}{Sch{\"o}lkopf\xfnm[ B]}.
\newblock \bibinfo{title}{Moûsai: Text-to-music generation with long-context latent diffusion}.
\newblock \bibinfo{journal}{arXiv preprint arXiv:230111757} \bibinfo{year}{2023};.
\bibitem[{Kahng et~al.(2018)Kahng, Thorat, Chau, Vi{\'e}gas and Wattenberg}]{ganlab}
\bibinfo{author}{Kahng\xfnm[ M]}, \bibinfo{author}{Thorat\xfnm[ N]}, \bibinfo{author}{Chau\xfnm[ DH]}, \bibinfo{author}{Vi{\'e}gas\xfnm[ FB]}, \bibinfo{author}{Wattenberg\xfnm[ M]}.
\newblock \bibinfo{title}{{GAN Lab}: Understanding complex deep generative models using interactive visual experimentation}.
\newblock \bibinfo{journal}{IEEE Transactions on Visualization and Computer Graphics} \bibinfo{year}{2018};\bibinfo{volume}{25}(\bibinfo{number}{1}):\bibinfo{pages}{310--320}.
\bibitem[{Lee et~al.(2023)Lee, Hoover, Strobelt, Wang, Peng, Wright et~al.}]{lee2023diffusion}
\bibinfo{author}{Lee\xfnm[ S]}, \bibinfo{author}{Hoover\xfnm[ B]}, \bibinfo{author}{Strobelt\xfnm[ H]}, \bibinfo{author}{Wang\xfnm[ ZJ]}, \bibinfo{author}{Peng\xfnm[ S]}, \bibinfo{author}{Wright\xfnm[ A]}, et~al.
\newblock \bibinfo{title}{Diffusion explainer: Visual explanation for text-to-image stable diffusion}.
\newblock \bibinfo{journal}{arXiv preprint arXiv:230503509} \bibinfo{year}{2023};.
\bibitem[{Park et~al.(2024)Park, Ju and Lee}]{park2024explaining}
\bibinfo{author}{Park\xfnm[ JH]}, \bibinfo{author}{Ju\xfnm[ YJ]}, \bibinfo{author}{Lee\xfnm[ SW]}.
\newblock \bibinfo{title}{Explaining generative diffusion models via visual analysis for interpretable decision-making process}.
\newblock \bibinfo{journal}{Expert Systems with Applications} \bibinfo{year}{2024};:\bibinfo{pages}{123231}.
\bibitem[{Liu et~al.(2021{\natexlab{a}})Liu, Cao, Su and Meng}]{diffsvc}
\bibinfo{author}{Liu\xfnm[ S]}, \bibinfo{author}{Cao\xfnm[ Y]}, \bibinfo{author}{Su\xfnm[ D]}, \bibinfo{author}{Meng\xfnm[ H]}.
\newblock \bibinfo{title}{{DiffSVC}: {A} diffusion probabilistic model for singing voice conversion}.
\newblock In: \bibinfo{booktitle}{Automatic Speech Recognition and Understanding Workshop}. \bibinfo{publisher}{{IEEE}}; \bibinfo{year}{2021}{\natexlab{a}}, p. \bibinfo{pages}{741--748}.
\bibitem[{Zhang et~al.(2023{\natexlab{b}})Zhang, Gu, Chen, Fang, Zou, Xue et~al.}]{zhang2023leveraging}
\bibinfo{author}{Zhang\xfnm[ X]}, \bibinfo{author}{Gu\xfnm[ Y]}, \bibinfo{author}{Chen\xfnm[ H]}, \bibinfo{author}{Fang\xfnm[ Z]}, \bibinfo{author}{Zou\xfnm[ L]}, \bibinfo{author}{Xue\xfnm[ L]}, et~al.
\newblock \bibinfo{title}{Leveraging content-based features from multiple acoustic models for singing voice conversion}.
\newblock \bibinfo{journal}{Machine Learning for Audio Workshop, Neural Information Processing Systems} \bibinfo{year}{2023}{\natexlab{b}};.
\bibitem[{Lu et~al.(2024)Lu, Ye, Xue, Tan, Liu and Guo}]{lu2024comosvc}
\bibinfo{author}{Lu\xfnm[ Y]}, \bibinfo{author}{Ye\xfnm[ Z]}, \bibinfo{author}{Xue\xfnm[ W]}, \bibinfo{author}{Tan\xfnm[ X]}, \bibinfo{author}{Liu\xfnm[ Q]}, \bibinfo{author}{Guo\xfnm[ Y]}.
\newblock \bibinfo{title}{Comosvc: Consistency model-based singing voice conversion}.
\newblock \bibinfo{journal}{arXiv preprint arXiv:240101792} \bibinfo{year}{2024};.
\bibitem[{Goodfellow et~al.(2020)Goodfellow, Pouget-Abadie, Mirza, Xu, Warde-Farley, Ozair et~al.}]{goodfellow2020generative}
\bibinfo{author}{Goodfellow\xfnm[ I]}, \bibinfo{author}{Pouget-Abadie\xfnm[ J]}, \bibinfo{author}{Mirza\xfnm[ M]}, \bibinfo{author}{Xu\xfnm[ B]}, \bibinfo{author}{Warde-Farley\xfnm[ D]}, \bibinfo{author}{Ozair\xfnm[ S]}, et~al.
\newblock \bibinfo{title}{Generative adversarial networks}.
\newblock \bibinfo{journal}{Communications of the ACM} \bibinfo{year}{2020};\bibinfo{volume}{63}(\bibinfo{number}{11}):\bibinfo{pages}{139--144}.
\bibitem[{Kingma and Welling(2014)}]{kingma2013vae}
\bibinfo{author}{Kingma\xfnm[ DP]}, \bibinfo{author}{Welling\xfnm[ M]}.
\newblock \bibinfo{title}{Auto-encoding variational bayes}.
\newblock In: \bibinfo{editor}{Bengio\xfnm[ Y]}, \bibinfo{editor}{LeCun\xfnm[ Y]}, editors. \bibinfo{booktitle}{International Conference on Learning Representations}. \bibinfo{year}{2014},.
\bibitem[{Sergios~Karagiannakos(2022)}]{karagiannakos2022diffusionmodels}
\bibinfo{author}{Sergios~Karagiannakos\xfnm[ NA]}.
\newblock \bibinfo{title}{Diffusion models: toward state-of-the-art image generation}.
\newblock \bibinfo{howpublished}{\url{https://theaisummer.com/diffusion-models/}}; \bibinfo{year}{2022}.
\bibitem[{O'Connor(2024)}]{assemblyai_diffusion_2024}
\bibinfo{author}{O'Connor\xfnm[ R]}.
\newblock \bibinfo{title}{Diffusion models for machine learning: Introduction}.
\newblock \bibinfo{howpublished}{\url{https://www.assemblyai.com/blog/diffusion-models-for-machine-learning-introduction/}}; \bibinfo{year}{2024}.
\bibitem[{T{\"{u}}rk et~al.(2009)T{\"{u}}rk, B{\"{u}}y{\"{u}}k, Haznedaroglu and Arslan}]{parallel-svc-2009-HMM}
\bibinfo{author}{T{\"{u}}rk\xfnm[ O]}, \bibinfo{author}{B{\"{u}}y{\"{u}}k\xfnm[ O]}, \bibinfo{author}{Haznedaroglu\xfnm[ A]}, \bibinfo{author}{Arslan\xfnm[ LM]}.
\newblock \bibinfo{title}{Application of voice conversion for cross-language rap singing transformation}.
\newblock In: \bibinfo{booktitle}{International Conference on Acoustics, Speech and Signal Processing}. \bibinfo{year}{2009}, p. \bibinfo{pages}{3597--3600}.
\bibitem[{Kobayashi et~al.(2014)Kobayashi, Toda, Neubig, Sakti and Nakamura}]{parallel-toda-2014}
\bibinfo{author}{Kobayashi\xfnm[ K]}, \bibinfo{author}{Toda\xfnm[ T]}, \bibinfo{author}{Neubig\xfnm[ G]}, \bibinfo{author}{Sakti\xfnm[ S]}, \bibinfo{author}{Nakamura\xfnm[ S]}.
\newblock \bibinfo{title}{Statistical singing voice conversion with direct waveform modification based on the spectrum differential}.
\newblock In: \bibinfo{booktitle}{International Speech Communication Association}. \bibinfo{year}{2014}, p. \bibinfo{pages}{2514--2518}.
\bibitem[{Kobayashi et~al.(2015)Kobayashi, Toda, Neubig, Sakti and Nakamura}]{parallel-toda-2015}
\bibinfo{author}{Kobayashi\xfnm[ K]}, \bibinfo{author}{Toda\xfnm[ T]}, \bibinfo{author}{Neubig\xfnm[ G]}, \bibinfo{author}{Sakti\xfnm[ S]}, \bibinfo{author}{Nakamura\xfnm[ S]}.
\newblock \bibinfo{title}{Statistical singing voice conversion based on direct waveform modification with global variance}.
\newblock In: \bibinfo{booktitle}{International Speech Communication Association}. \bibinfo{year}{2015}, p. \bibinfo{pages}{2754--2758}.
\bibitem[{Nachmani and Wolf(2019)}]{non-parallel-svc-facebook}
\bibinfo{author}{Nachmani\xfnm[ E]}, \bibinfo{author}{Wolf\xfnm[ L]}.
\newblock \bibinfo{title}{Unsupervised singing voice conversion}.
\newblock In: \bibinfo{booktitle}{International Speech Communication Association}. \bibinfo{year}{2019}, p. \bibinfo{pages}{2583--2587}.
\bibitem[{Chen et~al.(2019)Chen, Chu, Guo and Xu}]{non-parallel-svc-chenxin}
\bibinfo{author}{Chen\xfnm[ X]}, \bibinfo{author}{Chu\xfnm[ W]}, \bibinfo{author}{Guo\xfnm[ J]}, \bibinfo{author}{Xu\xfnm[ N]}.
\newblock \bibinfo{title}{Singing voice conversion with non-parallel data}.
\newblock In: \bibinfo{booktitle}{Multimedia Information Processing and Retrieval}. \bibinfo{year}{2019}, p. \bibinfo{pages}{292--296}.
\bibitem[{Huang et~al.(2022)Huang, Yang, Hayashi and Toda}]{self-supervised-vc}
\bibinfo{author}{Huang\xfnm[ WC]}, \bibinfo{author}{Yang\xfnm[ SW]}, \bibinfo{author}{Hayashi\xfnm[ T]}, \bibinfo{author}{Toda\xfnm[ T]}.
\newblock \bibinfo{title}{A comparative study of self-supervised speech representation based voice conversion}.
\newblock \bibinfo{journal}{IEEE Journal of Selected Topics in Signal Processing} \bibinfo{year}{2022};\bibinfo{volume}{16}(\bibinfo{number}{6}):\bibinfo{pages}{1308--1318}.
\bibitem[{Liu et~al.(2021{\natexlab{b}})Liu, Cao, Hu, Su and Meng}]{fastsvc}
\bibinfo{author}{Liu\xfnm[ S]}, \bibinfo{author}{Cao\xfnm[ Y]}, \bibinfo{author}{Hu\xfnm[ N]}, \bibinfo{author}{Su\xfnm[ D]}, \bibinfo{author}{Meng\xfnm[ H]}.
\newblock \bibinfo{title}{{FastSVC}: Fast cross-domain singing voice conversion with feature-wise linear modulation}.
\newblock In: \bibinfo{booktitle}{International Conference on Multimedia and Expo}. \bibinfo{year}{2021}{\natexlab{b}}, p. \bibinfo{pages}{1--6}.
\bibitem[{Takahashi et~al.(2022)Takahashi, Singh and Mitsufuji}]{zero-shot-roboust-svc-bgm}
\bibinfo{author}{Takahashi\xfnm[ N]}, \bibinfo{author}{Singh\xfnm[ MK]}, \bibinfo{author}{Mitsufuji\xfnm[ Y]}.
\newblock \bibinfo{title}{Robust one-shot singing voice conversion}.
\newblock \bibinfo{journal}{arXiv} \bibinfo{year}{2022};\bibinfo{volume}{abs/2210.11096}.
\bibitem[{Luo et~al.(2020)Luo, Hsu, Agres and Herremans}]{svc-technique}
\bibinfo{author}{Luo\xfnm[ Y]}, \bibinfo{author}{Hsu\xfnm[ C]}, \bibinfo{author}{Agres\xfnm[ K]}, \bibinfo{author}{Herremans\xfnm[ D]}.
\newblock \bibinfo{title}{Singing voice conversion with disentangled representations of singer and vocal technique using variational autoencoders}.
\newblock In: \bibinfo{booktitle}{International Conference on Acoustics, Speech and Signal Processing}. \bibinfo{year}{2020}, p. \bibinfo{pages}{3277--3281}.
\bibitem[{{SVC-Develop-Team}(2023)}]{SoftvcVITS2023}
\bibinfo{author}{{SVC-Develop-Team}\xfnm[]}.
\newblock \bibinfo{title}{{{SoftSVC} VITS Singing Voice Conversion}}.
\newblock \bibinfo{howpublished}{\url{https://github.com/svc-develop-team/so-vits-svc}}; \bibinfo{year}{2023}.
\bibitem[{Popov et~al.(2022)Popov, Vovk, Gogoryan, Sadekova, Kudinov and Wei}]{diffvc}
\bibinfo{author}{Popov\xfnm[ V]}, \bibinfo{author}{Vovk\xfnm[ I]}, \bibinfo{author}{Gogoryan\xfnm[ V]}, \bibinfo{author}{Sadekova\xfnm[ T]}, \bibinfo{author}{Kudinov\xfnm[ MS]}, \bibinfo{author}{Wei\xfnm[ J]}.
\newblock \bibinfo{title}{Diffusion-based voice conversion with fast maximum likelihood sampling scheme}.
\newblock In: \bibinfo{booktitle}{International Conference on Learning Representations}. \bibinfo{year}{2022},.
\bibitem[{Choi et~al.(2023)Choi, Lee and Lee}]{diff-hiervc}
\bibinfo{author}{Choi\xfnm[ H]}, \bibinfo{author}{Lee\xfnm[ S]}, \bibinfo{author}{Lee\xfnm[ S]}.
\newblock \bibinfo{title}{{Diff-HierVC}: Diffusion-based hierarchical voice conversion with robust pitch generation and masked prior for zero-shot speaker adaptation}.
\newblock In: \bibinfo{booktitle}{International Speech Communication Association}. \bibinfo{year}{2023}, p. \bibinfo{pages}{2283--2287}.
\bibitem[{Wang et~al.(2022{\natexlab{a}})Wang, Ju, Tan, He, Wu, Bian et~al.}]{audit}
\bibinfo{author}{Wang\xfnm[ Y]}, \bibinfo{author}{Ju\xfnm[ Z]}, \bibinfo{author}{Tan\xfnm[ X]}, \bibinfo{author}{He\xfnm[ L]}, \bibinfo{author}{Wu\xfnm[ Z]}, \bibinfo{author}{Bian\xfnm[ J]}, et~al.
\newblock \bibinfo{title}{{AUDIT:} audio editing by following instructions with latent diffusion models}.
\newblock In: \bibinfo{booktitle}{Neural Information Processing Systems}. \bibinfo{year}{2022}{\natexlab{a}},.
\bibitem[{Arrieta et~al.(2020)Arrieta, D{\'i}az-Rodr{\'i}guez, Del~Ser, Bennetot, Tabik, Barbado et~al.}]{arrieta2020explainable}
\bibinfo{author}{Arrieta\xfnm[ AB]}, \bibinfo{author}{D{\'i}az-Rodr{\'i}guez\xfnm[ N]}, \bibinfo{author}{Del~Ser\xfnm[ J]}, \bibinfo{author}{Bennetot\xfnm[ A]}, \bibinfo{author}{Tabik\xfnm[ S]}, \bibinfo{author}{Barbado\xfnm[ A]}, et~al.
\newblock \bibinfo{title}{Explainable artificial intelligence ({XAI}): Concepts, taxonomies, opportunities and challenges toward responsible {AI}}.
\newblock \bibinfo{journal}{Information Fusion} \bibinfo{year}{2020};\bibinfo{volume}{58}:\bibinfo{pages}{82--115}.
\bibitem[{Hohman et~al.(2018)Hohman, Kahng, Pienta and Chau}]{hohman2018visual}
\bibinfo{author}{Hohman\xfnm[ F]}, \bibinfo{author}{Kahng\xfnm[ M]}, \bibinfo{author}{Pienta\xfnm[ R]}, \bibinfo{author}{Chau\xfnm[ DH]}.
\newblock \bibinfo{title}{Visual analytics in deep learning: An interrogative survey for the next frontiers}.
\newblock \bibinfo{journal}{IEEE Transactions on Visualization and Computer Graphics} \bibinfo{year}{2018};\bibinfo{volume}{25}(\bibinfo{number}{8}):\bibinfo{pages}{2674--2693}.
\bibitem[{Wang et~al.(2020)Wang, Turko, Shaikh, Park, Das, Hohman et~al.}]{wang2020cnn}
\bibinfo{author}{Wang\xfnm[ ZJ]}, \bibinfo{author}{Turko\xfnm[ R]}, \bibinfo{author}{Shaikh\xfnm[ O]}, \bibinfo{author}{Park\xfnm[ H]}, \bibinfo{author}{Das\xfnm[ N]}, \bibinfo{author}{Hohman\xfnm[ F]}, et~al.
\newblock \bibinfo{title}{{CNN Explainer}: learning convolutional neural networks with interactive visualization}.
\newblock \bibinfo{journal}{IEEE Transactions on Visualization and Computer Graphics} \bibinfo{year}{2020};\bibinfo{volume}{27}(\bibinfo{number}{2}):\bibinfo{pages}{1396--1406}.
\bibitem[{Strobelt et~al.(2017)Strobelt, Gehrmann, Pfister and Rush}]{LSTMVis}
\bibinfo{author}{Strobelt\xfnm[ H]}, \bibinfo{author}{Gehrmann\xfnm[ S]}, \bibinfo{author}{Pfister\xfnm[ H]}, \bibinfo{author}{Rush\xfnm[ AM]}.
\newblock \bibinfo{title}{{LSTMVis}: A tool for visual analysis of hidden state dynamics in recurrent neural networks}.
\newblock \bibinfo{journal}{IEEE Transactions on Visualization and Computer Graphics} \bibinfo{year}{2017};\bibinfo{volume}{24}(\bibinfo{number}{1}):\bibinfo{pages}{667--676}.
\bibitem[{Wang et~al.(2018{\natexlab{a}})Wang, Gou, Shen and Yang}]{DQNViz}
\bibinfo{author}{Wang\xfnm[ J]}, \bibinfo{author}{Gou\xfnm[ L]}, \bibinfo{author}{Shen\xfnm[ HW]}, \bibinfo{author}{Yang\xfnm[ H]}.
\newblock \bibinfo{title}{{DQNViz}: A visual analytics approach to understand deep q-networks}.
\newblock \bibinfo{journal}{IEEE Transactions on Visualization and Computer Graphics} \bibinfo{year}{2018}{\natexlab{a}};\bibinfo{volume}{25}(\bibinfo{number}{1}):\bibinfo{pages}{288--298}.
\bibitem[{Wang et~al.(2021)Wang, He, Jin, Yang, Wang and Qu}]{M2lens}
\bibinfo{author}{Wang\xfnm[ X]}, \bibinfo{author}{He\xfnm[ J]}, \bibinfo{author}{Jin\xfnm[ Z]}, \bibinfo{author}{Yang\xfnm[ M]}, \bibinfo{author}{Wang\xfnm[ Y]}, \bibinfo{author}{Qu\xfnm[ H]}.
\newblock \bibinfo{title}{{M2Lens}: Visualizing and explaining multimodal models for sentiment analysis}.
\newblock \bibinfo{journal}{IEEE Transactions on Visualization and Computer Graphics} \bibinfo{year}{2021};\bibinfo{volume}{28}(\bibinfo{number}{1}):\bibinfo{pages}{802--812}.
\bibitem[{Liu et~al.(2016)Liu, Shi, Li, Li, Zhu and Liu}]{CNNVis}
\bibinfo{author}{Liu\xfnm[ M]}, \bibinfo{author}{Shi\xfnm[ J]}, \bibinfo{author}{Li\xfnm[ Z]}, \bibinfo{author}{Li\xfnm[ C]}, \bibinfo{author}{Zhu\xfnm[ J]}, \bibinfo{author}{Liu\xfnm[ S]}.
\newblock \bibinfo{title}{Towards better analysis of deep convolutional neural networks}.
\newblock \bibinfo{journal}{IEEE Transactions on Visualization and Computer Graphics} \bibinfo{year}{2016};\bibinfo{volume}{23}(\bibinfo{number}{1}):\bibinfo{pages}{91--100}.
\bibitem[{Yeh et~al.(2023)Yeh, Chen, Wu, Chen, Vi{\'e}gas and Wattenberg}]{AttentionViz}
\bibinfo{author}{Yeh\xfnm[ C]}, \bibinfo{author}{Chen\xfnm[ Y]}, \bibinfo{author}{Wu\xfnm[ A]}, \bibinfo{author}{Chen\xfnm[ C]}, \bibinfo{author}{Vi{\'e}gas\xfnm[ F]}, \bibinfo{author}{Wattenberg\xfnm[ M]}.
\newblock \bibinfo{title}{{AttentionViz}: A global view of transformer attention}.
\newblock \bibinfo{journal}{arXiv preprint arXiv:230503210} \bibinfo{year}{2023};.
\bibitem[{Norton and Qi(2017)}]{norton2017adversarial}
\bibinfo{author}{Norton\xfnm[ AP]}, \bibinfo{author}{Qi\xfnm[ Y]}.
\newblock \bibinfo{title}{Adversarial-playground: A visualization suite showing how adversarial examples fool deep learning}.
\newblock In: \bibinfo{booktitle}{IEEE Symposium on Visualization for Cyber Security}. \bibinfo{year}{2017}, p. \bibinfo{pages}{1--4}.
\bibitem[{Wang et~al.(2018{\natexlab{b}})Wang, Gou, Yang and Shen}]{GANViz}
\bibinfo{author}{Wang\xfnm[ J]}, \bibinfo{author}{Gou\xfnm[ L]}, \bibinfo{author}{Yang\xfnm[ H]}, \bibinfo{author}{Shen\xfnm[ HW]}.
\newblock \bibinfo{title}{{GANViz}: A visual analytics approach to understand the adversarial game}.
\newblock \bibinfo{journal}{IEEE Transactions on Visualization and Computer Graphics} \bibinfo{year}{2018}{\natexlab{b}};\bibinfo{volume}{24}(\bibinfo{number}{6}):\bibinfo{pages}{1905--1917}.
\bibitem[{Wang et~al.(2022{\natexlab{b}})Wang, Huang, Chandak, Zitnik and Gehlenborg}]{DrugExplorer}
\bibinfo{author}{Wang\xfnm[ Q]}, \bibinfo{author}{Huang\xfnm[ K]}, \bibinfo{author}{Chandak\xfnm[ P]}, \bibinfo{author}{Zitnik\xfnm[ M]}, \bibinfo{author}{Gehlenborg\xfnm[ N]}.
\newblock \bibinfo{title}{Extending the nested model for user-centric xai: A design study on gnn-based drug repurposing}.
\newblock \bibinfo{journal}{IEEE Transactions on Visualization and Computer Graphics} \bibinfo{year}{2022}{\natexlab{b}};\bibinfo{volume}{29}(\bibinfo{number}{1}):\bibinfo{pages}{1266--1276}.
\bibitem[{Huang et~al.(2023{\natexlab{b}})Huang, Violeta, Liu, Shi, Yasuda and Toda}]{Huang2023TheSV}
\bibinfo{author}{Huang\xfnm[ WC]}, \bibinfo{author}{Violeta\xfnm[ LP]}, \bibinfo{author}{Liu\xfnm[ S]}, \bibinfo{author}{Shi\xfnm[ J]}, \bibinfo{author}{Yasuda\xfnm[ Y]}, \bibinfo{author}{Toda\xfnm[ T]}.
\newblock \bibinfo{title}{The singing voice conversion challenge 2023}.
\newblock In: \bibinfo{booktitle}{Automatic Speech Recognition and Understanding Workshop}. \bibinfo{year}{2023}{\natexlab{b}}, p. \bibinfo{pages}{1--8}.
\bibitem[{Zhang et~al.(2023{\natexlab{c}})Zhang, Xue, Wang, Gu, Chen, Fang et~al.}]{amphion}
\bibinfo{author}{Zhang\xfnm[ X]}, \bibinfo{author}{Xue\xfnm[ L]}, \bibinfo{author}{Wang\xfnm[ Y]}, \bibinfo{author}{Gu\xfnm[ Y]}, \bibinfo{author}{Chen\xfnm[ X]}, \bibinfo{author}{Fang\xfnm[ Z]}, et~al.
\newblock \bibinfo{title}{Amphion: An open-source audio, music and speech generation toolkit}.
\newblock \bibinfo{journal}{arXiv} \bibinfo{year}{2023}{\natexlab{c}};\bibinfo{volume}{abs/2312.09911}.
\bibitem[{van~den Oord et~al.(2016)van~den Oord, Dieleman, Zen, Simonyan, Vinyals, Graves et~al.}]{Wavenet}
\bibinfo{author}{van~den Oord\xfnm[ A]}, \bibinfo{author}{Dieleman\xfnm[ S]}, \bibinfo{author}{Zen\xfnm[ H]}, \bibinfo{author}{Simonyan\xfnm[ K]}, \bibinfo{author}{Vinyals\xfnm[ O]}, \bibinfo{author}{Graves\xfnm[ A]}, et~al.
\newblock \bibinfo{title}{{WaveNet}: {A} generative model for raw audio}.
\newblock In: \bibinfo{booktitle}{Speech Synthesis Workshop}. \bibinfo{publisher}{{ISCA}}; \bibinfo{year}{2016}, p. \bibinfo{pages}{125}.
\bibitem[{Ho et~al.(2020)Ho, Jain and Abbeel}]{ddpm}
\bibinfo{author}{Ho\xfnm[ J]}, \bibinfo{author}{Jain\xfnm[ A]}, \bibinfo{author}{Abbeel\xfnm[ P]}.
\newblock \bibinfo{title}{Denoising diffusion probabilistic models}.
\newblock \bibinfo{journal}{Neural Information Processing Systems} \bibinfo{year}{2020};\bibinfo{volume}{33}:\bibinfo{pages}{6840--6851}.
\bibitem[{Radford et~al.(2023)Radford, Kim, Xu, Brockman, McLeavey and Sutskever}]{whisper}
\bibinfo{author}{Radford\xfnm[ A]}, \bibinfo{author}{Kim\xfnm[ JW]}, \bibinfo{author}{Xu\xfnm[ T]}, \bibinfo{author}{Brockman\xfnm[ G]}, \bibinfo{author}{McLeavey\xfnm[ C]}, \bibinfo{author}{Sutskever\xfnm[ I]}.
\newblock \bibinfo{title}{Robust speech recognition via large-scale weak supervision}.
\newblock In: \bibinfo{booktitle}{International Conference on Machine Learning}. \bibinfo{organization}{PMLR}; \bibinfo{year}{2023}, p. \bibinfo{pages}{28492--28518}.
\bibitem[{Qian et~al.(2022)Qian, Zhang, Gao, Ni, Lai, Cox et~al.}]{contentvec}
\bibinfo{author}{Qian\xfnm[ K]}, \bibinfo{author}{Zhang\xfnm[ Y]}, \bibinfo{author}{Gao\xfnm[ H]}, \bibinfo{author}{Ni\xfnm[ J]}, \bibinfo{author}{Lai\xfnm[ CI]}, \bibinfo{author}{Cox\xfnm[ D]}, et~al.
\newblock \bibinfo{title}{{ContentVec}: An improved self-supervised speech representation by disentangling speakers}.
\newblock In: \bibinfo{booktitle}{International Conference on Machine Learning}. \bibinfo{organization}{PMLR}; \bibinfo{year}{2022}, p. \bibinfo{pages}{18003--18017}.
\bibitem[{Jadoul et~al.(2018)Jadoul, Thompson and de~Boer}]{parselmouth}
\bibinfo{author}{Jadoul\xfnm[ Y]}, \bibinfo{author}{Thompson\xfnm[ B]}, \bibinfo{author}{de~Boer\xfnm[ B]}.
\newblock \bibinfo{title}{Introducing {P}arselmouth: A {P}ython interface to {P}raat}.
\newblock \bibinfo{journal}{Journal of Phonetics} \bibinfo{year}{2018};\bibinfo{volume}{71}:\bibinfo{pages}{1--15}.
\bibitem[{Zhao et~al.(2023)Zhao, Wang, Guo, Lu and Chen}]{zhao2023contextwing}
\bibinfo{author}{Zhao\xfnm[ Y]}, \bibinfo{author}{Wang\xfnm[ X]}, \bibinfo{author}{Guo\xfnm[ C]}, \bibinfo{author}{Lu\xfnm[ M]}, \bibinfo{author}{Chen\xfnm[ S]}.
\newblock \bibinfo{title}{Contextwing: Pair-wise visual comparison for evolving sequential patterns of contexts in social media data streams}.
\newblock \bibinfo{journal}{Proceedings of the ACM on Human-Computer Interaction} \bibinfo{year}{2023};\bibinfo{volume}{7}(\bibinfo{number}{CSCW1}):\bibinfo{pages}{1--31}.
\bibitem[{Rossi et~al.(2010)Rossi, Lipsey and Freeman}]{rossi2018evaluation}
\bibinfo{author}{Rossi\xfnm[ PH]}, \bibinfo{author}{Lipsey\xfnm[ MW]}, \bibinfo{author}{Freeman\xfnm[ HE]}.
\newblock \bibinfo{title}{Evaluation: A systematic approach}.
\newblock \bibinfo{journal}{Canadian Journal of University Continuing Education} \bibinfo{year}{2010};\bibinfo{volume}{36}(\bibinfo{number}{2}).
\bibitem[{Wang et~al.(2022{\natexlab{c}})Wang, Wang, Zhu, Wu, Li, Xue et~al.}]{opencpop}
\bibinfo{author}{Wang\xfnm[ Y]}, \bibinfo{author}{Wang\xfnm[ X]}, \bibinfo{author}{Zhu\xfnm[ P]}, \bibinfo{author}{Wu\xfnm[ J]}, \bibinfo{author}{Li\xfnm[ H]}, \bibinfo{author}{Xue\xfnm[ H]}, et~al.
\newblock \bibinfo{title}{{Opencpop}: {A} high-quality open source chinese popular song corpus for singing voice synthesis}.
\newblock In: \bibinfo{booktitle}{International Speech Communication Association}. \bibinfo{publisher}{{ISCA}}; \bibinfo{year}{2022}{\natexlab{c}}, p. \bibinfo{pages}{4242--4246}.
\bibitem[{Yamagishi et~al.(2019)Yamagishi, Veaux, MacDonald et~al.}]{vctk}
\bibinfo{author}{Yamagishi\xfnm[ J]}, \bibinfo{author}{Veaux\xfnm[ C]}, \bibinfo{author}{MacDonald\xfnm[ K]}, et~al.
\newblock \bibinfo{title}{{CSTR VCTK Corpus}: English multi-speaker corpus for cstr voice cloning toolkit (version 0.92)}.
\newblock \bibinfo{journal}{University of Edinburgh The Centre for Speech Technology Research} \bibinfo{year}{2019};.
\bibitem[{Huang et~al.(2021)Huang, Chen, Ren, Liu, Cui and Zhao}]{multisinger}
\bibinfo{author}{Huang\xfnm[ R]}, \bibinfo{author}{Chen\xfnm[ F]}, \bibinfo{author}{Ren\xfnm[ Y]}, \bibinfo{author}{Liu\xfnm[ J]}, \bibinfo{author}{Cui\xfnm[ C]}, \bibinfo{author}{Zhao\xfnm[ Z]}.
\newblock \bibinfo{title}{{Multi-Singer}: Fast multi-singer singing voice vocoder with {A} large-scale corpus}.
\newblock In: \bibinfo{booktitle}{ACM International Conference on Multimedia}. \bibinfo{publisher}{{ACM}}; \bibinfo{year}{2021}, p. \bibinfo{pages}{3945--3954}.
\bibitem[{Zhang et~al.(2022)Zhang, Li, Wang, Deng, Liu, Ren et~al.}]{m4singer}
\bibinfo{author}{Zhang\xfnm[ L]}, \bibinfo{author}{Li\xfnm[ R]}, \bibinfo{author}{Wang\xfnm[ S]}, \bibinfo{author}{Deng\xfnm[ L]}, \bibinfo{author}{Liu\xfnm[ J]}, \bibinfo{author}{Ren\xfnm[ Y]}, et~al.
\newblock \bibinfo{title}{M4singer: {A} multi-style, multi-singer and musical score provided mandarin singing corpus}.
\newblock In: \bibinfo{booktitle}{Neural Information Processing Systems}. \bibinfo{year}{2022},.

\end{thebibliography}


\clearpage
\appendix

\section{Metrics Definition}
\label{app:metircs}
\begin{itemize}
    
    
    \item  \textbf{Singer Similarity (Dembed)} quantitatively assesses the similarity between the timbre of the original singer's voice and the converted voice. It's calculated using the cosine similarity between feature vectors representing the timbre characteristics of the two voices. A higher similarity score indicates more timbre similarity. 
    
    \item \textbf{F0 Pearson Correlation Coefficient (F0CORR)}  measures the Pearson Correlation Coefficient between the F0 values of the converted singing voice and the target voice. It assesses the linear relationship between the F0 contours of the two voices. A higher F0CORR indicates a stronger correlation and better F0 similarity.
    
    \item \textbf{Fréchet Audio Distance (FAD)} is a reference-free evaluation metric to evaluate the quality of audio samples. FAD correlates more closely with human perception. A lower FAD score indicates a higher quality of the audio.
    
    \item  \textbf{F0 Root Mean Square Error (F0RMSE)} measures the Root Mean Square Error of the Fundamental Frequency (F0) values between the converted singing voice and the target voice. It quantifies how accurately the F0 of the converted voice matches that of the target voice. A lower F0RMSE indicates better F0 accuracy.
    
    \item  \textbf{Mel-cepstral Distortion (MCD)} assesses the quality of the generated speech by comparing the discrepancy between generated and ground-truth singing voice. It measures how different the two sequences of Mel cepstra are. A lower MCD indicates better quality.
 
\end{itemize}

\section{Dataset}
\label{app:dataset}

We use five datasets for our diffusion-based SVC model training: Opencpop \cite{opencpop}, SVCC training data \cite{Huang2023TheSV}, VCTK \cite{vctk}, OpenSinger \cite{multisinger}, and M4Singer \cite{m4singer}. In total, these datasets contain 83.1 hours of speech and 87.2 hours of singing data. The mapping between the singer name defined in the dataset and the singer ID displayed in the SingVisio system is listed in Table~\ref{table:mapping_of_singer}. The mapping between the song name defined in the dataset and song ID in the SingVisio system is listed in Table~\ref{table:mapping_of_song}.

\begin{table}[H]
\caption{Mapping of singer name and singer ID}
\label{table:mapping_of_singer}
\centering
\scalebox{0.85}{
\begin{tabular}{|c|c|c|c|}
\hline
\textbf{Dataset} & \textbf{Singer Name} & \textbf{Gender} & \textbf{Singer ID} \\ \hline
\multirow{6}{*}{\textbf{SVCC}} & SF1 & Female & Singer 1 \\
 & SM1  & Male  & Singer 2 \\
 & CDF1 & Female & Singer 3 \\
 & CDM1 & Male & Singer 4 \\
 & IDF1 & Female & Singer 5 \\
 & IDM1 & Male & Singer 6 \\ \cline{1-4}
\multirow{7}{*}{\textbf{M4Singer}} & Alto-1 & Female & Singer 7 \\
 & Alto-7 & Female & Singer 8 \\
 & Bass-1 & Male & Singer 9 \\
 & Soprano-2 & Female & Singer 10 \\
 & Tenor-5 & Male & Singer 11 \\
 & Tenor-6 & Male & Singer 12 \\
 & Tenor-7 & Male & Singer 13 \\ \cline{1-4}
\textbf{Opencpop} & Opencpop & Female & Singer 14 \\ \hline
\end{tabular}
}
\end{table}

\begin{table}[H]
\caption{Mapping of song name and song ID.}
\label{table:mapping_of_song}
\centering
\scalebox{0.85}{
\begin{tabularx}{0.55\textwidth}{|m{1.4cm}|m{2cm}|m{1.2cm}|>{\centering\arraybackslash}X|}
\hline
\textbf{Dataset} & \textbf{Utterance ID} & \textbf{Song ID} & \textbf{Lyrics} \\ \hline
\multirow{13}{*}[-10ex]{\textbf{SVCC}} & 30001 & Song 1 & Hey Jude, don't make it bad. \\ \cline{2-4}
 & 30002 & Song 2 & Take a sad song and make it better. \\ \cline{2-4}
 & 30003 & Song 3 & Remember to let her into your heart. \\ \cline{2-4}
 & 10001 & Song 4 & Everything is fine. \\ \cline{2-4}
 & 10030 & Song 5 & Were you lying all the time? \\ \cline{2-4}
 & 10120 & Song 6 & Now, I need  \\ \cline{2-4}
 & 10140 & Song 7 & Hey, I love you. \\ \cline{2-4}
 & 30005 & Song 15 & You know that its fool who plays it cool. \\ \cline{2-4}
 & 30006 & Song 16 & Na, na, na, na, na, na, na, na, na, na, na, na, hey, Jude. \\ \cline{2-4}
 & 30009 & Song 17 & When they all should let us be. \\ \cline{2-4}
 & 30016 & Song 18 &  Let it be. Let it be. Let it be.  \\ \cline{2-4}
 & 30022 & Song 19 & Take my breath away \\ \cline{2-4}
 & 30019 & Song 20 & Watching every motion In my foolish lover's game \\ \hline
 \multirow{10}{*}[-12ex]{\textbf{M4Singer}} &  Alto-1\_\begin{CJK*}{UTF8}{gbsn}美错 \end{CJK*}\_0014 & Song 8 & \begin{CJK*}{UTF8}{gbsn}  美丽的错误往往最接近真实 \end{CJK*} \\ \cline{2-4}
 & Bass-1\_\begin{CJK*}{UTF8}{gbsn}十年 \end{CJK*}\_0008 & Song 9 & \begin{CJK*}{UTF8}{gbsn}  陪在一个陌生人左右 \end{CJK*} \\ \cline{2-4}
 & Soprano-2\_\begin{CJK*}{UTF8}{gbsn}同桌的你 \end{CJK*}\_0018 & Song 10 & \begin{CJK*}{UTF8}{gbsn}  谁遇见多愁善感的你 \end{CJK*} \\ \cline{2-4}
 & Tenor-5\_\begin{CJK*}{UTF8}{gbsn}爱笑的眼睛 \end{CJK*}\_0010 & Song 11  & \begin{CJK*}{UTF8}{gbsn}  这爱的城市虽然拥挤\end{CJK*}  \\ \cline{2-4}
 & Alto-7\_\begin{CJK*}{UTF8}{gbsn}寂寞沙洲冷 \end{CJK*}\_0000 & Song 12  & \begin{CJK*}{UTF8}{gbsn} \multirow{2}{*}{\shortstack{河畔的风放肆拼命的吹,\\无端拨弄离人的眼泪}} \end{CJK*} \\ 
 & Tenor-6\_\begin{CJK*}{UTF8}{gbsn}寂寞沙洲冷 \end{CJK*}\_0002 & Song 12  &  \\ \cline{2-4}
 & Alto-7\_\begin{CJK*}{UTF8}{gbsn}寂寞沙洲冷 \end{CJK*}\_0011 & Song 13  & \begin{CJK*}{UTF8}{gbsn}  \multirow{4}{*}{\shortstack{当记忆的线缠绕过往支离\\破碎, 是慌乱占据了心扉}}  \end{CJK*}  \\
 & Tenor-7\_\begin{CJK*}{UTF8}{gbsn}寂寞沙洲冷 \end{CJK*}\_0013 & Song 13  &  \\ 
 & Tenor-6\_\begin{CJK*}{UTF8}{gbsn}寂寞沙洲冷 \end{CJK*}\_0020 & Song 13  &   \\ 
 & Bass-1\_\begin{CJK*}{UTF8}{gbsn}寂寞沙洲冷 \end{CJK*}\_0021 & Song 14  & \\ \hline
\end{tabularx}
}
\end{table}

\begin{figure*}[th!]
    \centering
    \includegraphics[width=0.8\textwidth]{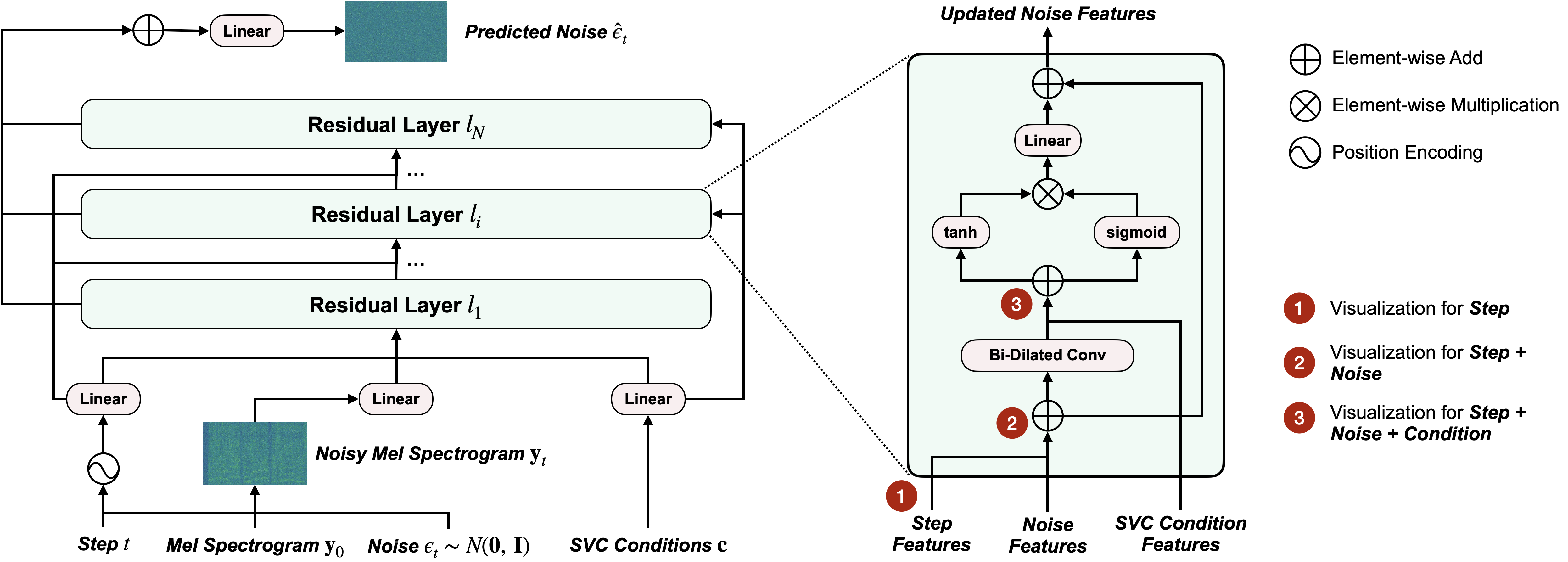}
    \caption{The architecture of DiffWaveNetSVC~\cite{amphion,zhang2023leveraging}. We select the \textit{\textbf{Step}}, \textit{\textbf{Step+Noise}}, and \textit{\textbf{Step+Noise+Condition}} to project and visualize in SingVisio's Project View (Section~\ref{sec:projection-view})}
    \label{fig:diffwavenetsvc}
\end{figure*}

\section{Architecture of Diffusion-based SVC}
\label{app:svc_model}
The architecture of DiffWaveNetSVC is shown in Fig.~\ref{fig:diffwavenetsvc}.



\clearpage
\clearpage
\begin{minipage}{\textwidth}
    \centering
    \includegraphics[page=1,width=\textwidth]{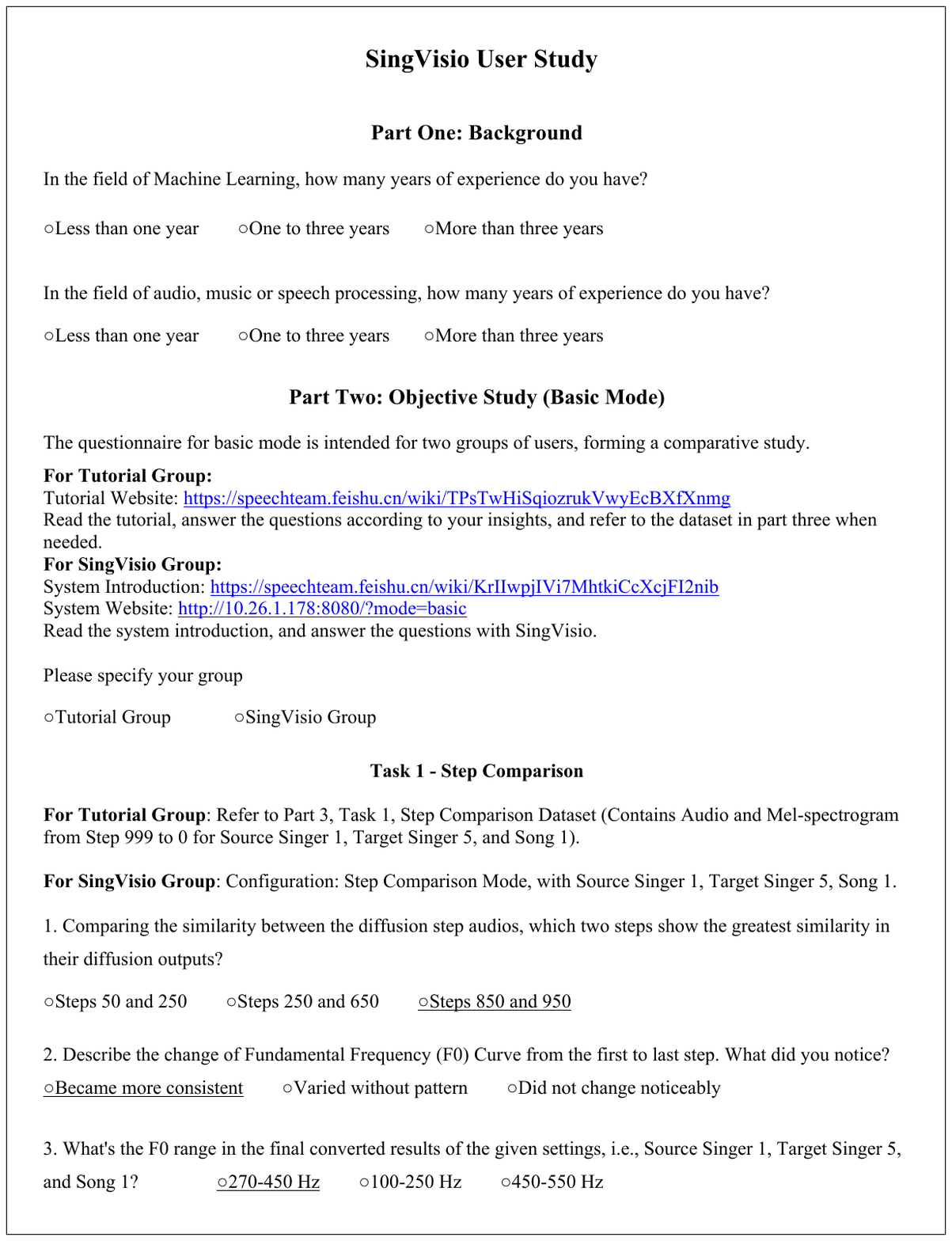}
    \captionof{figure}{SingVisio User Study 1/7}
\end{minipage}
\clearpage
\begin{minipage}{\textwidth}
    \centering
    \includegraphics[page=2,width=\textwidth]{figs/user_study.pdf}
    \captionof{figure}{SingVisio User Study 2/7}
\end{minipage}
\clearpage
\begin{minipage}{\textwidth}
    \centering
    \includegraphics[page=3,width=\textwidth]{figs/user_study.pdf}
    \captionof{figure}{SingVisio User Study 3/7}
\end{minipage}
\clearpage
\begin{minipage}{\textwidth}
    \centering
    \includegraphics[page=4,width=\textwidth]{figs/user_study.pdf}
    \captionof{figure}{SingVisio User Study 4/7}
\end{minipage}
\clearpage
\begin{minipage}{\textwidth}
    \centering
    \includegraphics[page=5,width=\textwidth]{figs/user_study.pdf}
    \captionof{figure}{SingVisio User Study 5/7}
\end{minipage}
\clearpage
\begin{minipage}{\textwidth}
    \centering
    \includegraphics[page=6,width=\textwidth]{figs/user_study.pdf}
    \captionof{figure}{SingVisio User Study 6/7}
\end{minipage}
\clearpage
\begin{minipage}{\textwidth}
    \centering
    \includegraphics[page=7,width=\textwidth]{figs/user_study.pdf}
    \captionof{figure}{SingVisio User Study 7/7}
\end{minipage}
\clearpage

\end{document}